\documentclass[a4paper, 11pt]{article}

\usepackage{amsthm}
\usepackage{amsmath}
\usepackage{amssymb}
\usepackage{graphicx}
\usepackage[round]{natbib}
\usepackage{bbm}
\usepackage{textcomp}
\usepackage{url}
\usepackage{color}
\usepackage{framed}
\usepackage{lscape}
\usepackage{multirow}
\usepackage{booktabs}
\usepackage{setspace}
\usepackage[margin = 2.5cm]{geometry}
\definecolor{shadecolor}{gray}{0.9}

\theoremstyle{plain}  %default 
\newtheorem{thm}{Theorem}[section] 
 
\newtheorem{prop}[thm]{Proposition}

\theoremstyle{definition} 
\newtheorem{defn}{Definition}[section]

\newtheorem{assump}{Assumption}[section]

\theoremstyle{remark}

\newcommand{\diff}{\,\mathrm{d}}

\newcommand{\ES}{\operatorname{ES}}

\newcommand{\R}{\mathbb{R}}

\newcommand{\F}{\mathcal{F}}
\newcommand{\A}{\mathsf{A}}

\newcommand{\one}{\mathbbm{1}}

\newcommand{\VaR}{\operatorname{VaR}}

\begin{document}

\title{Murphy Diagrams: Forecast Evaluation of Expected Shortfall\thanks{We thank seminar and conference participants in Heidelberg and Augsburg (Statistische Woche 2016) for helpful comments. Johanna Ziegel gratefully acknowledges financial support of the Swiss National Science Foundation. The work of Fabian Kr\"uger and Alexander Jordan has been funded by
the European Union Seventh Framework Programme under grant agreement
290976. They also thank the Klaus Tschira Foundation for
infrastructural support at the Heidelberg Institute for Theoretical
Studies (HITS). The opinions expressed in this article are those of the authors do not necessarily reflect the views of Raiffeisen Schweiz.}}
\author{ Johanna F.~Ziegel\\University of Bern \and Fabian Kr\"uger \\ Heidelberg University \and Alexander Jordan \\ Heidelberg Institute for \\ Theoretical Studies \and Fernando Fasciati \\Raiffeisen Schweiz}

\maketitle

\abstract{Motivated by the Basel 3 regulations, recent studies have considered joint forecasts of Value-at-Risk and Expected Shortfall. A large family of scoring functions can be used to evaluate forecast performance in this context. However, little intuitive or empirical guidance is currently available, which renders the choice of scoring function awkward in practice. We therefore develop graphical checks (Murphy diagrams) of whether one forecast method dominates another under a relevant class of scoring functions, and propose an associated hypothesis test. We illustrate these tools with simulation examples and an empirical analysis of S\&P 500 and DAX returns.\\
 
}

\section{Introduction}

The Basel 3 standard on minimum capital requirements for market risk \citep{Basel2016} uses Expected Shortfall (ES), rather than Value-at-Risk (VaR), to quantify the risk of a bank's portfolio. As described by \citet[Chapter 8]{McneilEtAl2015}, ES possesses several desirable theoretical properties. However, it also has a major drawback: It is not elicitable, i.e.~there is no scoring function that sets the incentive to report ES honestly, or that can be used to compare ES forecasts' accuracy.\footnote{As detailed below, a scoring function (or loss function) assigns a real-valued score, given a forecast and a realizing observation.} As a partial remedy to this problem, \citet[henceforth FZ]{FisslerZiegel2015} show that ES is jointly elicitable with VaR and characterize the class of scoring functions that can be used to evaluate forecasts of type (VaR, ES). \citet{FisslerEtAl2015} provide a nontechnical introduction and discuss regulatory implications.\\

In applied work, it is challenging to select a specific member function from the FZ family on either economic or statistical grounds. Motivated by this problem, we present a mixture representation using elementary members of the FZ family, which is mathematically similar to recent results by \citet{EhmEtAl2015} for quantiles and expectiles. The mixture representation gives rise to Murphy diagrams which allow to check whether one forecast dominates another under a relevant class of scoring functions.\footnote{The name of the diagrams alludes to the meteorologist Allan H.~Murphy (1931--1997) who pioneered similar diagrams in the context of a binary dependent variable (see \citealt{Murphy1977}, as well as \citealt[][p.~519]{EhmEtAl2015}).}	 While this class could be the entire FZ family, we argue that a subfamily which emphasizes ES -- rather than VaR -- is economically more plausible in the light of the Basel 3 standard. Analyzing the robustness of forecast rankings across this class of scoring functions is relevant both conceptually and practically, and referred to as \emph{forecast dominance} in the following. \\ 

Forecast dominance holds at the population level - that is, it is defined in terms of expected performance, which is unobservable. Statistical tests are designed to detect significant deviations of the observed performance from hypotheses about expected performance; see e.g.~\cite{Diebold1995} and  \cite{ClarkMcCracken2013}. In the present context, such tests are complicated by the fact that the null hypothesis refers to performance under all elementary members of the mixture representation, i.e.~on a grid of parameters. Following a suggestion by \citet[Section 3.4]{EhmEtAl2015}, we discuss a permutation test which accounts for this circumstance via multiple testing corrections. The labels of two forecasting methods are randomly switched to enforce the null hypothesis of equal forecast performance, allowing the computation of $p$-values by Monte Carlo simulation. While a formal investigation of this test is left for future research, simulation evidence points to satisfactory size and power properties.\\

In an empirical case study, we evaluate forecasts for daily log returns of the S\&P 500 and DAX stock market indices. Three models with varying degree of sophistication are considered: the HEAVY model \citep{ShephardSheppard2010} with access to the past's intra-daily data competes against two models using merely end-of-day data, a GARCH(1,1) model \citep{Bollerslev1986} and a naive `historical simulation' model. Our results suggest that the HEAVY model tends to outperform its competitors, as indicated by Murphy diagrams and tests of forecast dominance.\\

We emphasize that our interest lies in \textit{comparative} forecast evaluation -- that is, we seek to compare the (VaR, ES) forecasts of two competing methods.\footnote{In financial jargon, the word `backtesting' is sometimes used as a synonym for `forecast evaluation'.} Comparative evaluation is important to select a suitable forecasting method in practice, especially given the wealth of data sources and statistical techniques that could plausibly be used to generate forecasts. Comparative forecast evaluation is different from \textit{absolute} evaluation which aims to determine whether a given forecast method possesses certain desirable optimality properties. The Basel 2 procedure of counting VaR `violations', i.e.~the number of times the actual return fell below the VaR forecast, is an example of absolute forecast evaluation. See \cite{NoldeZiegel2016} for a detailed discussion of comparative versus absolute evaluation of financial forecasts.\\

The contributions of the present paper include a mixture representation of the FZ family in Section \ref{sec:2}, which yields the Murphy diagrams, and a permutation test for the hypothesis of forecast dominance in Section \ref{sec:3}. We identify a class of scoring functions primarily suited for the evaluation of the expected shortfall component in a forecast of type ($\VaR, \ES$), illustrate its use in an empirical case study in Section \ref{sec:case}, and draw a link to European put options in Section \ref{sec:options}. A discussion in Section 6 concludes.\\

\section{Consistent Scoring Functions for VaR and Expected Shortfall}\label{sec:2}

To keep notation light, we start with a single-period outcome and move on to time-series considerations in the next section. Let $Y \in \mathbb{R}$ be a random variable describing the single-period return of a financial asset, where a negative return, $Y < 0$, corresponds to a loss. Value-at-Risk (VaR) and Expected Shortfall (ES) are popular measures of tail risk. Let $F$ denote the distribution of $Y$, and assume that $Y$ has finite mean. Then for a given level $\alpha \in (0,1)$, the VaR and ES are defined as
\[
\VaR_{\alpha}(F) = \inf\{z \in \R : F(z) \ge \alpha\}
\]
and
\[
\ES_{\alpha}(F) = \frac{1}{\alpha}\int_0^{\alpha} \VaR_u(F) \diff u.
\]
We are interested in small values of $\alpha$, in particular $\alpha=0.025$ which is the level that the \citet{Basel2016} requests for ES predictions. Then, $\VaR_\alpha$ and $\ES_\alpha$ will typically have negative values. Our sign convention corresponds to the sign convention of utility functions as used in \citet{Delbaen2012} and it implies that $\VaR_\alpha \ge \ES_\alpha$ always holds.\\

Following \cite{Gneiting2011}, \cite{EhmEtAl2015}, \cite{Patton2016} and others, it is now widely recognized that consistent scoring functions are essential for comparing point forecasts. Consistency implies that, on average, a misspecified model may not outperform a correct model. As discussed in \cite{FisslerZiegel2015}, $\ES_{\alpha}$ cannot be evaluated consistently without joint consideration of $\VaR_\alpha$, so we stack the two functionals to obtain the two-dimensional functional $$\mathrm{T}_\alpha(F) = (\VaR_\alpha(F),\ES_\alpha(F))'.$$
As return distributions we consider members of the class $\F_1$ of distributions with finite mean and unique quantiles. The latter assumption allows us to simplify our presentation and does not seem restrictive in the context of financial returns. For example, the HEAVY and GARCH models used in our case study (Section 4) clearly satisfy the assumption. As forecasts of type $\mathrm{T}_\alpha$ we consider elements of the action domain $\A_0 = \{x \in \R^2 : x_1 \ge x_2 \}$, thereby ruling out irrational forecasts that violate $\VaR_\alpha \ge \ES_\alpha$. The following definition formalizes the notion of a consistent scoring function for $\mathrm{T}_{\alpha}$.

\begin{defn}
A \emph{scoring function} $\mathrm{S}:\A_0\times \R \to \R$ is a function such that $\int \mathrm{S}(x,y) \diff F(y)$ exists for all $F \in \F_1$, $x \in \A_0$. The scoring function $\mathrm{S}$ is called \emph{consistent} for $\mathrm{T}_{\alpha}$ if
\begin{equation}\label{eq:argmin}
\mathbb{E}(\mathrm{S}(\mathrm{T}_{\alpha}(F),Y)) \le \mathbb{E}(\mathrm{S}(x,Y))
\end{equation}
for all $x\in \A_0$ and all random variables $Y$ with distribution in $\F_1$. The scoring function $\mathrm{S}$ is \emph{strictly consistent} if equality in \eqref{eq:argmin} implies $x = \mathrm{T}_{\alpha}(F)$. 
\end{defn}

Equation (\ref{eq:argmin}) says that, in expectation, it is a forecaster's best possible action to state the forecast $\mathrm{T}_{\alpha}(F),$ rather than an arbitrary alternative $x \in \A_0$. In this sense, a consistent scoring function sets the incentive for honest and accurate forecasting of $\mathrm{T}_{\alpha}$. Importantly, there is not only one scoring function that is consistent for $\mathrm{T}_{\alpha}$. Instead, there is a whole family of scoring functions with this property.\footnote{ 
The situation is  similar for other functionals, i.e., there is typically a whole family of scoring functions that are consistent for a given functional. For example, \cite{Savage1971} identifies a family of scoring functions that are consistent for the mean, and \cite{Gneiting2011b} describes the family of scoring functions that are consistent for a quantile.} As shown by \citet[Section 5] {FisslerZiegel2015}, consistent scoring functions for $\mathrm{T}_{\alpha}$ take the form $\mathrm{S}(x_1,x_2,y),$ where $x_1$ is a forecast of $\VaR_\alpha$, $x_2$ is a forecast of $\ES_\alpha$, and $y$ is the realization. Here, we consider normalized scores for which $\mathrm{S}(y,y,y)=0$ holds true. This normalization is in line with much of the existing literature \citep[e.g.][]{Gneiting2011}; other normalizations can easily be accommodated. Corollary 5.5 of \cite{FisslerZiegel2015} implies that all scoring functions $\mathrm{S}$ of the form 
\begin{equation}\label{eq:SVaRES}
\begin{split}
\mathrm{S}(x_1,x_2,y) &= \big(\one\{y\le x_1\} - \alpha \big)\big(G_1(x_1) - G_1(y)\big)\\ 
&\quad+ G_2(x_2)\Big(\frac{1}{\alpha}\one\{y\le x_1\}(x_1 - y) - (x_1 - x_2)\Big) \\ 
&\quad - \big(\mathcal G_2(x_2) - \mathcal{G}_2(y)\big), 
\end{split}
\end{equation}
are consistent scoring functions for $\mathrm{T}_\alpha$, where $G_1$, $G_2,$ and $\mathcal{G}_2$ are functions from $\mathbb{R}$ to $\mathbb{R}$, $\mathcal{G}_2'= G_2$, $G_1$ and $G_2$ are increasing, $G_2 \ge 0$ and $\int G_1(y)\diff F(y)$, $\int \mathcal{G}_2(y)\diff F(y)$ exist and are finite for all $F \in \F_1$. If $G_2$ is strictly increasing, we obtain strict consistency. For example, the choice $G_1(z) = 0, G_2(z) = \exp(z)/(1+\exp(z))$ satisfies all of these requirements but there are many alternatives. Subject to regularity conditions, all normalized consistent scoring functions on the action domain $\A_0$ are of the form \eqref{eq:SVaRES}.\\

\citet{Patton2016} and others have demonstrated that the choice of scoring function is relevant for the ranking of two competing forecasts in the presence of model misspecification and non-nested information sets, both of which are common in practice. Here we seek to develop methods for comparing forecasts under a class of scoring functions, thus avoiding the need to select a single specific function. We therefore make the following definition of forecast dominance which is analogous to \citet[Definition 1]{EhmEtAl2015}.
\begin{defn}\label{def:dominance}
Let $\alpha \in (0,1)$ and let $\mathcal{S}$ be a class of consistent scoring functions for $\mathrm{T}_\alpha$. For two (possibly random) forecasts $(X_1^A,X_2^A)$ and $(X_1^B,X_2^B)$ made by methods A and B, respectively, we say that method A \emph{weakly dominates} method $B$ \emph{with respect to $\mathcal{S}$} if 
\[
\mathbb{E}\left(\mathrm{S}(X_1^A, X_2^A,Y)\right) \le \mathbb{E}\left(\mathrm{S}(X_1^B,X_2^B,Y)\right) , \quad \text{for all $\mathrm{S} \in \mathcal{S}$,}
\]
where the expectations are with respect to the joint distribution of $(X_1^A, X_2^A, X_1^B, X_2^B, Y)$.
\end{defn}
When $\mathcal{S}$ is a `small' class it feasible to check an empirical version of dominance for all members. And importantly, once dominance has been established for a given class it can be translated to the extension including all mixtures, e.g.~dominance with respect to $\{S_1, S_2\}$ implies dominance with respect to $\{a S_1 + b S_2 : a, b\ge 0\}$. This simple observation is the basis for so-called Murphy diagrams which are graphical tools to check for forecast dominance empirically with respect to all consistent scoring functions \citep{EhmEtAl2015}. To this end, \citet{EhmEtAl2015} provide mixture representations of the families of consistent scoring functions for quantiles and expectiles. In order to derive similar methodology for $\mathrm{T}_\alpha$, the following result presents a mixture representation for consistent scoring functions of the form given in \eqref{eq:SVaRES}.

\begin{prop}\label{prop:mix}
Let $\alpha \in (0,1)$. For $v_1, v_2,y \in \R$, $(x_1,x_2) \in \A_0$, we define
\begin{align*}
\mathrm{S}_{v_1}(x_1,y) &= (\one\{y\le x_1\}-\alpha)\big(\one\{v_1 \le x_1\} - \one\{v_1 \le y\}\big)\\ 
\mathrm{S}_{v_2}(x_1,x_2,y)&= \one\{v_2 \le x_2\}\left(\frac{1}{\alpha}  \one\{y\le x_1\}(x_1 - y) - (x_1 - v_2)\right) + \one\{v_2 \le y\}(y-v_2).
\end{align*}
Let $H_1$ be a locally finite measure and $H_2$ a measure that is finite on all intervals of the form $(-\infty,x]$, $x \in \R$. Then all scoring functions $\mathrm{S}:\A_0 \times \R \to \R$ that are of the form \eqref{eq:SVaRES} can be written as
\begin{equation}\label{eq:mix}
\mathrm{S}(x_1,x_2,y) = \int \mathrm{S}_{v_1}(x_1,y) \diff H_1(v_1) + \int \mathrm{S}_{v_2}(x_1,x_2,y) \diff H_2(v_2).
\end{equation}
The scores at \eqref{eq:mix} are consistent for $\mathrm{T}_\alpha$. They are strictly consistent if $H_2$ puts positive mass on all open intervals.
\end{prop}

The \emph{elementary scores} $\mathrm{S}_{v_1}$ and $\mathrm{S}_{v_2}$ are themselves consistent scoring functions for $\mathrm{T}_\alpha$, which follows immediately by choosing Dirac-measures for $H_1$ or $H_2$ in \eqref{eq:mix}. Note that $\mathrm{S}_{v_1}$ for $\mathrm{T}_\alpha$ is also the elementary score in the class of consistent scoring functions for $\alpha$-quantiles as identified by \citet{EhmEtAl2015}, an unsurprising result given that $\VaR_\alpha$ is an $\alpha$-quantile. The elementary score $\mathrm{S}_{v_1}(x_1,y)$ goes to zero as $v_1 \rightarrow \pm \infty$. The second elementary score for $\mathrm{T}_\alpha$, $\mathrm{S}_{v_2}(x_1, x_2,y)$, takes a more complex form in that it depends on the joint forecast $(x_1, x_2)'$ and the realization $y$. It goes to zero as $v_2 \rightarrow + \infty$, and converges to $(1/\alpha)(\one\{y \le x_1\}-\alpha)(x_1-y)$ as $v_2 \rightarrow -\infty$. This explains the different restrictions on the corresponding mixing measures $H_1$ and $H_2$ in Proposition \ref{prop:mix}.\\

We now identify a subclass of consistent scoring functions for $\mathrm{T}_\alpha$ whose members emphasize the evaluation of the $\ES_\alpha$ component. The first integral in (\ref{eq:mix}) corresponds to the mixture representation of consistent scoring functions for quantiles \citep[Theorem 1a]{EhmEtAl2015}, a class that in our context only evaluates the $\VaR_\alpha$ forecast and ignores $\ES_\alpha$. Hence, choosing anything but a constant $H_1$ puts unnecessary emphasis on the $\VaR_\alpha$ component of a forecast of type $\mathrm{T}_\alpha$. The second integral corresponds to the evaluation of $\ES_{\alpha}$, conditional on $\VaR_{\alpha}$, where we cannot completely extinguish $\VaR_\alpha$ in the evaluation due to the results on the (non-)elicitability of $\ES_\alpha$. Hence, we define $\mathcal{S}_2$ as the class of all consistent scoring functions for $\mathrm{T}_\alpha$ as given at \eqref{eq:mix} with a constant $H_1$ (such that the first integral is zero), and focus on this class in the following.\\

Our focus on $\mathcal{S}_2$ is motivated by the aim to maximize the impact of the $\ES_\alpha$ component in evaluation, which is in line with the emphasis set in Basel 3. Focusing on $\mathcal{S}_2$ also seems justified from a statistical perspective: First, $\mathcal{S}_2$ contains positively homogeneous scoring functions for $\mathrm{T}_\alpha$ for all possible degrees of homogeneity; see \citet[][Section 2.3.1 and Theorem 6]{NoldeZiegel2016}. As discussed there, positively homogeneous scoring functions enjoy a number of attractive properties. Second, \citet{DimitriadisBayer2017} investigate several members of $\mathcal{S}_2$ in a regression framework. They argue that moving beyond $\mathcal{S}_2$ (i.e., considering non-constant choices of $H_1$ in Equation \ref{prop:mix}) does not improve the numerical performance of their estimators.\\

The mixture representation at \eqref{eq:mix} allows graphical displays of the performance of $\mathrm{T}_\alpha$ forecasts with respect to the elementary scores of $\mathcal{S}_2$,
\[
v_2 \mapsto \mathbb{E} (\mathrm{S}_{v_2}(X_1, X_2, Y)),
\]
where the expectation is with respect to the joint distribution of $(X_1, X_2, Y)$. In practice, the expectation is estimated by the average observed score. Examples of these displays, called Murphy diagrams \citep{EhmEtAl2015}, are given in Figure \ref{fig:murphy} in Section \ref{sec:case}. The diagrams provide simple graphical checks of whether one forecast dominates another under all scoring functions in $\mathcal{S}_2$. Specifically, Proposition \ref{prop:mix} implies that the forecast of method A dominates that of method B with respect to $\mathcal{S}_2$ if and only if
\[
\mathbb{E}\left(\mathrm{S}_{v_2}(X_1^A,X_2^A,Y)\right) \le  \mathbb{E}\left(\mathrm{S}_{v_2}(X_1^B,X_2^B,Y)\right) \quad\text{for all $v_2\in \mathbb{R}$};
\]
compare \citet[][Corollary 1]{EhmEtAl2015}.\\

Clearly, one could also consider forecast dominance for $\mathrm{T}_\alpha$ with respect to all consistent scoring functions. The procedures described in the following can be adapted to this case; an extension that is conceptually simple yet tedious in practice. This is because one needs to check inequalities across two grids of parameters, $v_1$ and $v_2$. Instead, when focusing on $\mathcal{S}_2$, it suffices to check inequalities along a single grid for $v_2$. We give results for all consistent scoring functions as a robustness check in Appendix D.

\section{Testing forecast dominance}\label{sec:3}

Here we first translate the methodology from Section \ref{sec:2} into a time series context, and then introduce a test of forecast dominance based on the elementary scores.

\subsection{Comparing time series forecasts}

So far, we have only considered a one-period forecasting problem. In most financial applications, however, the goal is to predict a time series $\{Y_t\}_{t \in \mathbb{N}}$, such as a sequence of asset returns observed at trading days $t = 1, 2, \ldots$. Furthermore, let $X_t = (X_{t, 1}, X_{t, 2})' \in \A_0$ denote the $(\VaR_\alpha, \ES_\alpha)$ forecast of $Y_t$, with the understanding that $X_t$ is based on an appropriate information set $\mathcal{W}_{t-1}$ generated by data available at time $t- 1$. In applications, we seek to make forecasts and realizations comparable across time. We therefore require the following assumption.
\begin{assump}\label{as:station}
The time series $\{Z_t\}_{t \in \mathbb{N}}$ with $Z_t = (X_t, Y_t)' \in \A_0\times\mathbb{R}$ is stationary and ergodic, with stationary distribution $F_Z$. 
\end{assump}

This assumption rules out deterministic time trends, structural breaks and seasonalities, among others. At the same time, the forecasts and realizations are allowed to fluctuate over time, as long as the fluctuations `wash out' eventually. In particular, many multivariate autoregressive models \citep[e.g.][]{Luetkepohl2005} or stochastic volatility models \citep[e.g.][]{HarveyEtAl1994} are stationary.\\

Consider any consistent scoring function $\mathrm{S}$ for $\mathrm{T}_\alpha$. Assumption \ref{as:station} implies that the distribution of the random variable $\mathrm{S}(X_t, Y_t)$ does not depend on time, $t$. In particular, this holds when $\mathrm{S}$ equals an elementary score $\mathrm{S}_{v_2}$ from Proposition \ref{prop:mix}. We can thus define the notion of an expected elementary score, as follows. 
Consider a sequence of forecasts $\{X_t\}_{t \in \mathbb{N}}$ and corresponding realizations $\{Y_t\}_{t \in \mathbb{N}}$ which jointly define a stationary time series as in Assumption \ref{as:station}. The expected elementary scores for this process are given by
\begin{equation}\label{eq:expect}
\mathbb{E}\left(\mathrm{S}_{v_2}(X_{t,1}, X_{t, 2}, Y_t)\right) = \int_{\A_0\times\mathbb{R}}  \mathrm{S}_{v_2}(x_1,x_2,y)~ \diff F_Z(x_1,x_2,y),
\end{equation}
where $F_Z$ is defined in Assumption \ref{as:station}. Based on this definition, a notion of forecast dominance `on average over time' follows naturally: 

\begin{defn}\label{def:dom}
Let $\{X_t^A\}_{t \in \mathbb{N}}$ and $\{X_t^B\}_{t \in \mathbb{N}}$ denote two competing sequences of forecasts of $\mathrm{T}_\alpha$, and let $\{Y_t\}_{t \in \mathbb{N}}$ denote the corresponding realizations such that $\{(X_t^A,Y_t)\}_{t \in \mathbb{N}}$ and $\{(X_t^B,Y_t)\}_{t \in \mathbb{N}}$ both satisfy Assumption \ref{as:station} with stationary distributions $F_{Z}^A$ and $F_{Z}^B$, respectively. We say that method A \emph{weakly dominates} method B with respect to $\mathcal{S}_2$ if
\[
%\mathbb{E}\left(S_{v_1}(X_{t,1}^A, Y_t)\right) & \le  & \mathbb{E}\left(S_{v_1}(X_{t,1}^B, Y_t)\right) ~\forall~v_1, \\ 
\mathbb{E}\left(\mathrm{S}_{v_2}(X_{t,1}^A, X_{t, 2}^A, Y_t)\right)  \le  \mathbb{E}\left(\mathrm{S}_{v_2}(X_{t,1}^B, X_{t, 2}^B, Y_t)\right) \quad\text{for all $v_2 \in \R$},
\]
where the expectations are as at \eqref{eq:expect} with respect to the corresponding stationary distribution.
\end{defn}

Under standard regularity conditions, the expectations in Definition \ref{def:dom} can be consistently estimated by empirical averages over observed forecasts and realizations at dates $t = 1, \ldots, T$, e.g.~as $T \rightarrow \infty$ it holds that $$\frac{1}{T} \sum_{t=1}^T \mathrm{S}_{v_2}(X_{t,1}^A, X_{t, 2}^A, Y_t)~\stackrel{a.s.}{\rightarrow}~\mathbb{E}\left(\mathrm{S}_{v_2}(X_{t,1}^A, X_{t, 2}^A, Y_t)\right),$$
and analogously for method B.

\subsection{Testing for forecast dominance}\label{sec:testing}

We are interested in the following null hypothesis: 
\begin{center}
\textit{H0: Method A weakly dominates method B}; 
\end{center}
Definition \ref{def:dom} gives a formal statement of the hypothesis. The test procedure, which we detail in Appendix B, can be summarized as follows:
\begin{itemize}
\item \textbf{Stage 1:} Test the null hypothesis that methods A and B perform equally well under a given elementary score, against the one-sided alternative that B performs strictly better, i.e.~for a given $v_2$ the pointwise null and alternative hypotheses are
\begin{align*}
&H0_{v_2}: \mathbb{E}\left(\mathrm{S}_{v_2}(X_{t,1}^A, X_{t, 2}^A, Y_t)\right)  =  \mathbb{E}\left(\mathrm{S}_{v_2}(X_{t,1}^B, X_{t, 2}^B, Y_t)\right), \\
&H1_{v_2}: \mathbb{E}\left(\mathrm{S}_{v_2}(X_{t,1}^A, X_{t, 2}^A, Y_t)\right)  >  \mathbb{E}\left(\mathrm{S}_{v_2}(X_{t,1}^B, X_{t, 2}^B, Y_t)\right).
\end{align*} This test is repeated for threshold values $v_2$ on a predefined grid, yielding a sequence of pointwise $p$-values.
\item \textbf{Stage 2:} Compute corrected $p$-values which are designed to control the family-wise error rate (FWER) of the procedure. The FWER is defined as the probability of making at least one false rejection, i.e.~rejecting $H0_{v_2}$ for at least one grid point $v_2$ at which A does not perform worse than B. Reject H0 if the minimum of the corrected $p$-values is below the chosen significance level.
\end{itemize}\vspace{0.2cm}

The tests in the first stage are one-sided $t$-tests of the null hypothesis that the expected score difference between models A and B is zero. To implement the correction in the second stage, we apply the \cite{WestfallYoung1993} algorithm to the pointwise $p$-values; see also \cite{CoxLee2008} who investigate the properties of the algorithm in the context of functional data, i.e., the null hypotheses refer to a grid of values of some parameter as in our case. The most important implementation choice is how to simulate $p$-values under the null hypothesis that model A weakly dominates model B, as formulated in Definition 3.1. Our approach draws an i.i.d.~sample where each realization of joint pointwise $p$-values is simulated by reassigning the forecasts' labels with equal probability at each time step. This enforces equality of the expected elementary scores of models A and B for all $v_2$, thus representing the boundary of the null hypothesis of {weak} dominance. We call the minimum of the corrected $p$-values the \emph{minimal Westfall-Young} $p$-value.\\

Our results rest on the assumption that the score differences are independent over time (henceforth, IOT). Given that we consider one-day ahead forecasts, this assumption follows standard practice in the econometric forecasting literature, which is to consider autocorrelation up to lag $\tau - 1$, where $\tau$ is the forecast horizon; see e.g.~\citet[Section 7]{ClarkMcCracken2013}. Consistent with the IOT assumption, the tests in the first stage do not account for possible autocorrelation, and the label switches in the second stage are performed independently over time.\\

It is possible to implement the test differently taking into account correlation of score differences over time. We consider this alternative implementation in Appendix D where we apply the resulting tests to the data example of Section \ref{sec:case}. There, we also consider the use of both types of elementary scores for $\mathrm{T}_\alpha$, an extension where the pointwise null and alternative hypotheses are defined on two independent grids of values for $v_1$ and $v_2$, respectively. \\

We acknowledge that there are two open issues about the testing procedure just described. First, the relabeling step enforces \textit{exchangeability} of the two models' scores. While exchangeability implies equality of expected scores, the converse is not true. It is unclear whether this imbalance vanishes or remains asymptotically. Second, our testing procedure controls its size $\alpha$ at the boundary of the null hypothesis, for equality of the expected scores of models A and B. Intuitively,  and as conjectured by \citet[p.~522]{EhmEtAl2015}, one would expect that the test's rejection probability is smaller than $\alpha$ in the interior of the null hypothesis, when A strictly dominates B at some grid points. However, a formal proof of this intuition is beyond the scope of this paper.\\

In view of these open issues, we investigate the testing procedure by simulation.\footnote{We use the R programming language \citep{rcite} for all simulations and empirical analyses in this paper.} The data generating process is  similar to the HEAVY forecasting model which we use in the empirical analysis of Section \ref{sec:case}. To this end, we first create data from the deterministic process
\[
\sigma_t^2 = 0.5 ~\text{RK}_{t-1} + 0.7~\sigma^2_{t-1},
\]
where $\text{RK}_t$ is a 'realized kernel' measure of intra-day volatility \citep{Barndorff2008, Barndorff2009}, i.e.~between end-of-day time $t-1$ and $t$, and $\sigma_0^2 = 0.35$.  We use the RK values as recorded in the S\&P 500 data set in Section \ref{sec:case}, and assume that the return at day $t$ is given by $$R_t=\sqrt{\frac{\nu-2}{\nu}}~\sigma_t~X_t,$$ where $\{X_t\}_{t \in \mathbb{N}}$ is a sequence of i.i.d.~random variables which are $t$-distributed with $\nu = 6$ degrees of freedom.\footnote{The factor $\sqrt{(\nu-2)/\nu}$ accounts for the fact that the variance of a $t$-distributed variable equals $\nu/(\nu-2)$. Hence, the factor ensures that the conditional variance of $R_t$ is given by $\sigma_t^2$.} Given knowledge of the process and the sequence $\{\text{RK}_j\}_{j \le t-1}$, the perfect $\mathrm{T}_\alpha$ forecast for $R_t$ consists of
\begin{align*}
\VaR_{t|t-1, \alpha}^* &= \sqrt{\frac{\nu-2}{\nu}}~ \sigma_t~ q_{\alpha, \nu}, \\
\ES_{t|t-1, \alpha}^* &= \frac{1}{\alpha} \int_{0}^{\alpha} \VaR_{t|t-1, z}^* \diff z,
\end{align*}
where $q_{\alpha, \nu}$ is the $\alpha$-quantile of the $t$-distribution with $\nu$ degrees of freedom. We fix $\alpha=0.025$ and consider two forecasting models $m \in \{1, 2\}$, with forecasts given as follows:
\begin{align*}
\VaR_{t|t-1, m} &= \VaR_{t|t-1}^* + \varepsilon_{t, m}, \\
\ES_{t|t-1, m} &= \ES_{t|t-1}^* + \varepsilon_{t, m},
\end{align*}
where $\varepsilon_{t, m} \sim \mathcal{N}(0, \zeta_m),$ independently of $t$ and $m$. The variance term $\zeta_m \ge 0$ is a measure of expected deviation from optimality. The limiting case $\zeta_m = 0$ means that $\varepsilon_{t, m} = 0$ almost surely, and corresponds to perfect forecasts. Note that model $m$ incurs the same error in both components of its $\mathrm{T}_\alpha$ forecast. To investigate the size and power of the proposed testing procedure, we experiment with different choices of $\zeta_1$ and $\zeta_2$.\footnote{For all investigations, the testing procedure uses 50 equally spaced grid points for $v_2$, whose range is determined from the empirical range of all forecasts and realizations. Furthermore, we use $500$ iterations of the relabeling procedure in Stage 2 of each test.}\\

\begin{figure}
\centering
\begin{tabular}{cc}
Size & Power \\ [-0.5cm]
\includegraphics[width = 0.4\textwidth]{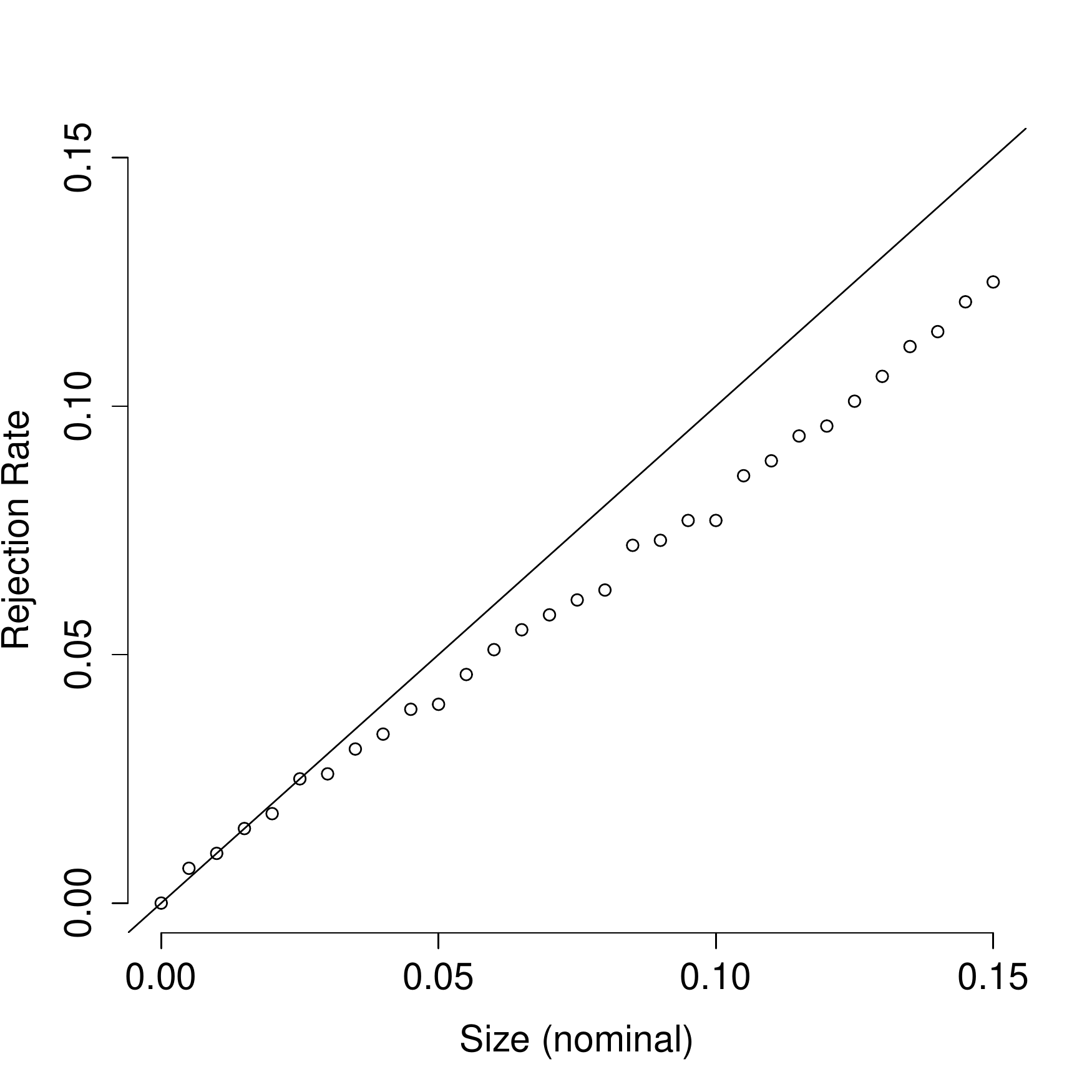} &
\includegraphics[width = 0.4\textwidth]{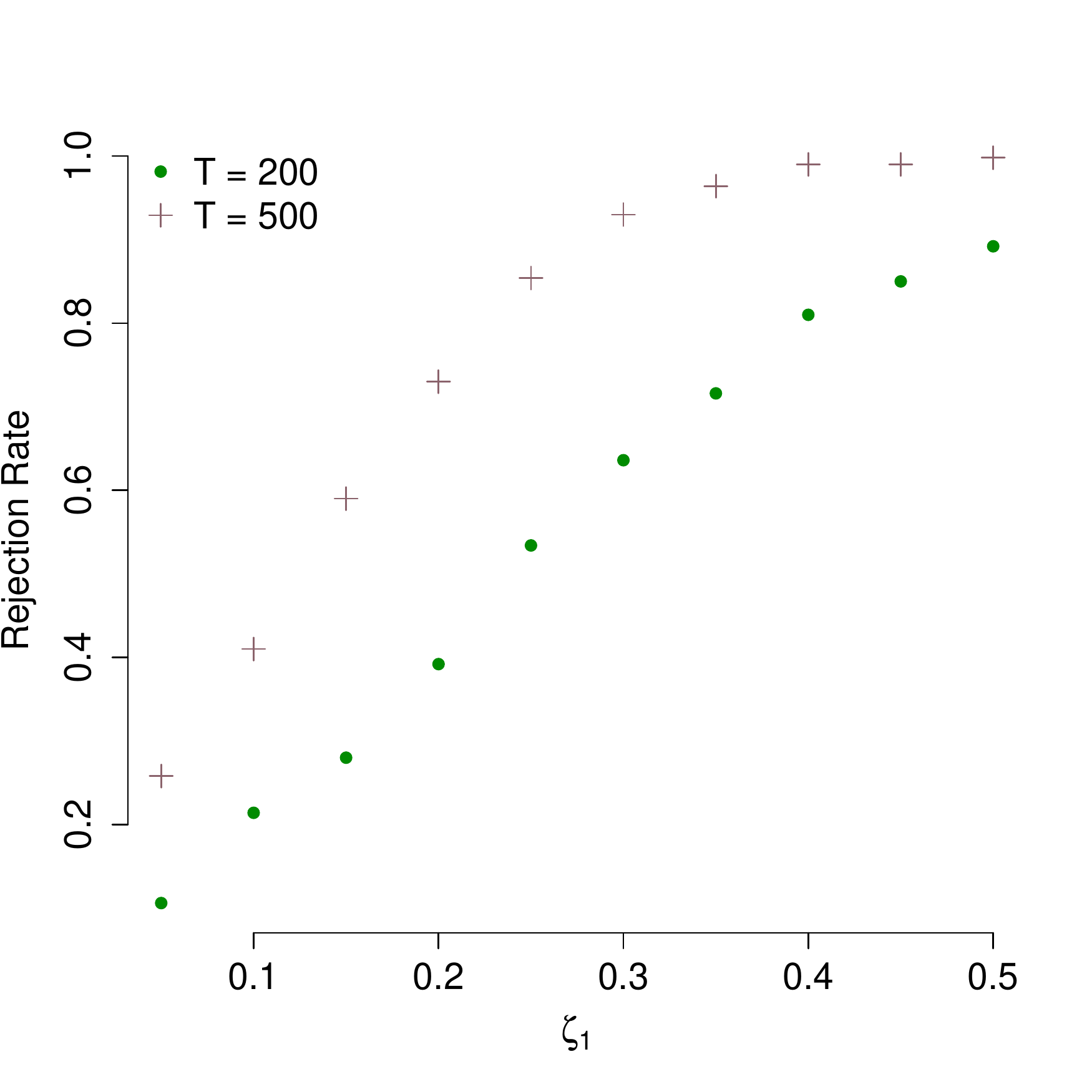} 
\end{tabular}
\caption{\textbf{Monte Carlo analysis of test for forecast dominance}. Left - Westfall-Young rejection rates plotted against nominal test size; data simulated under H0 ($\zeta_1 = \zeta_2 = 1$). Each simulated data set comprises $T = 500$ time periods. The results are based on $1000$ Monte Carlo iterations. Right - Rejection rates of Westfall-Young tests at level 5 percent; data simulated under the alternative ($\zeta_1 > 0, \zeta_2 = 0$) for two sample sizes ($T = 200$ and $T = 500$). The results are based on $500$ Monte Carlo iterations for each combination of $\zeta_1$ and $T$.}
\label{fig:wysim}
\end{figure}

First consider the case $\zeta_1 = \zeta_2 = 1$. This means that both models have equal expected scores, which is consistent with weak dominance. To investigate the size of our procedure, we simulate $1000$ data sets comprising forecasts and realizations for $T = 500$ time periods, yielding a sample of 1000 minimal Westfall-Young $p$-values. The left panel of Figure \ref{fig:wysim} illustrates the results for nominal levels ranging from 0 to 15 percent. We observe a conservative behavior of the test with empirical size slightly below the nominal level. These results seem  satisfactory, and suggest that our approach of controlling the test's size at the boundary of the null hypothesis is sufficient in the present context.\\

We further investigate the power behavior at the 5 percent level. Consider the case of $\zeta_1 > 0$ and $\zeta_2 = 0$, which means that the second model issues perfect forecasts while the first model deviates from optimality to a degree measured by $\zeta_1$. This implies a dominance relationship in favor of the second model following results from \cite{Tsyplakov2014}.\footnote{Both models have access to the same information base, which is used optimally by the second model, but suboptimally by the first model. \cite{Tsyplakov2014} shows that this setup implies dominance of the second model under all proper scoring rules.} In the simulation, we vary $\zeta_1$ from 0.05 to 0.5, consider $T=200, 500$, and generate a sample of 500 minimal Westfall-Young $p$-values for each combination of $\zeta_1$ and $T$. The right panel of Figure \ref{fig:wysim} suggests a monotonic power increase in both variables with convergence to 1. These qualitative features are in line with common sense, and thus provide a sanity check for the testing procedure. \\

\section{Empirical Results for S\&P 500 and DAX Returns}\label{sec:case}

In this section, we apply our methodology to compare forecasts for the returns of two stock indices, the S\&P 500 and the DAX. The return of the index (S\&P 500 or DAX) is defined as
$$R_t = 100 \times (\log P_t - \log P_{t-1}),$$
where $P_t$ is the level of the index at the end of trading day $t$. As before, let $\mathcal{W}_{t-1}$ denote the information set generated by data up to day $t-1$. We consider three models for  daily log returns with corresponding $(\VaR_\alpha, \ES_{\alpha})$ forecasts at level $\alpha = 0.025$:

\begin{itemize}
\item The HEAVY model \citep{ShephardSheppard2010} which uses intra-daily realized measures to model the time-varying variance of financial returns. The model posits that
\begin{equation}\label{heavy1}
\mathbb{V}(R_{t}|\mathcal{W}_{t-1}) = \sigma_{t}^2 = ~\omega + \gamma~\text{RK}_{t-1} + \beta~ \sigma_{t-1}^2, 
\end{equation}
where $\mathbb{V}$ denotes variance, and $\text{RK}_{t-1}$ is the realized kernel measure computed from intra-daily price movements at day $t-1$. The quantities $\omega >0, \gamma$ and $\beta$  are model parameters which we estimate via the quasi-likelihood method described in \citet[][Section 2.4.1]{ShephardSheppard2010}. We re-fit the model only on the first trading day of each month using a rolling window of $1500$ observations, i.e. roughly six years of daily data. We further assume that, conditional on $\mathcal{W}_{t-1}$, $(\sqrt{(\nu-2)/\nu}~\sigma_{t-1})^{-1}~R_{t}$ follows a $t$-distribution with six degrees of freedom. This set of assumptions yields an estimate of $\VaR_\alpha$ and $\ES_\alpha$ of $R_{t}$, conditional on $\mathcal{W}_{t-1}$.
\item A GARCH(1,1) model as proposed by \cite{Bollerslev1986}. The variance specification coincides with Equation (\ref{heavy1}), except that the squared daily return, $R_{t-1}^2$, is used in place of $\text{RK}_{t-1}$. As for the HEAVY model, we assume a $t$-distribution with six degrees of freedom for the scaled conditional return distribution.
\item The empirical unconditional $\VaR_\alpha$ and $\ES_\alpha$ computed from the returns in the $1500$ observations up until day $t-1$. This approach resembles `historical simulation' (HS) methods which are popular in practice \citep[see e.g.][Section 9.2.3]{McneilEtAl2015}.
\end{itemize}

Our analysis is based on data from \url{http://realized.oxford-man.ox.ac.uk/}; this source covers both daily closing prices and realized measures computed from intra-daily data. We construct forecasts for the period from January 2006 to January 2016.\footnote{More precisely, the S\&P 500 sample comprises $2420$ observations from January 6, 2006 to January 25, 2016; the DAX sample comprises $2494$ observations from January 4, 2006 to January 25, 2016.} The entire analysis is out-of-sample, i.e.~we evaluate the forecasts against realizations which were not used for model fitting. \\

\begin{table}
	\centering
	\begin{tabular}{llccc}
		& & Avg.~$\VaR_\alpha$  & Avg.~$\ES_\alpha$ & $\VaR_\alpha$ `violation' rate \\ \toprule
		\multirow{3}{*}{S\&P 500\quad}  & HEAVY & -2.056 & -2.736 & 0.042 \\
& GARCH(1,1) & -2.184 & -2.906 & 0.040 \\& 
HS & -2.761 & -4.028 & 0.029 \\
 \midrule
		\multirow{3}{*}{DAX\quad}  & HEAVY & -2.500 & -3.326 & 0.037 \\& 
GARCH(1,1) & -2.607 & -3.469 & 0.036 \\& 
HS & -3.130 & -4.493 & 0.025 \\
 \bottomrule
	\end{tabular}
	\caption{\textbf{Summary statistics for empirical forecasts}. Sample period ranges from January 2006 to January 2016 (daily data). The $\VaR_\alpha$ `violation' rate is the fraction of days for which the actual returns falls below the $\VaR_\alpha$ forecast and should not exceed $\alpha = 0.025$.}
	\label{fc_sum}
\end{table}

Table \ref{fc_sum} presents summary statistics on the forecasts; Figure \ref{fig:forecasts} in Appendix \ref{app:figs} presents corresponding time series plots. On average, the HS model produces lower forecasts than those by the other two methods. For the S\&P 500 data set, the average $\VaR_\alpha$ forecast is $-2.056$ for HEAVY, compared to $-2.184$ for GARCH and $-2.761$ for HS. The violation rates of the $\VaR_\alpha$ forecasts are $4.2$ percent (HEAVY), $4$ percent (GARCH) and $2.9$ percent (HS), with all three methods exceeding the nominal level of $2.5$ percent, partially due to the negative returns surrounding the 2007-09 financial crisis. Figure \ref{fig:forecasts} in Appendix \ref{app:figs} shows that the HEAVY and GARCH forecasts are highly correlated, and display much more time variation than the forecasts of the simple HS method. The latter observation shows that the HEAVY and GARCH models are much quicker to react to changes in the market environment than the HS method.\\

\begin{figure}[p]
		\centering
		\begin{tabular}{cccc}
			\multicolumn{2}{c}{\textbf{\large S\&P 500}} & \multicolumn{2}{c}{\textbf{\large DAX}} \\[1ex] \toprule
			&&&\\[-1ex]
			\multicolumn{2}{c}{Murphy diagrams} & \multicolumn{2}{c}{Murphy diagrams} \\
			\multicolumn{2}{c}{(HEAVY, GARCH, HS)} & \multicolumn{2}{c}{(HEAVY, GARCH, HS)} \\
			\multicolumn{2}{c}{\includegraphics[width = .4\textwidth]{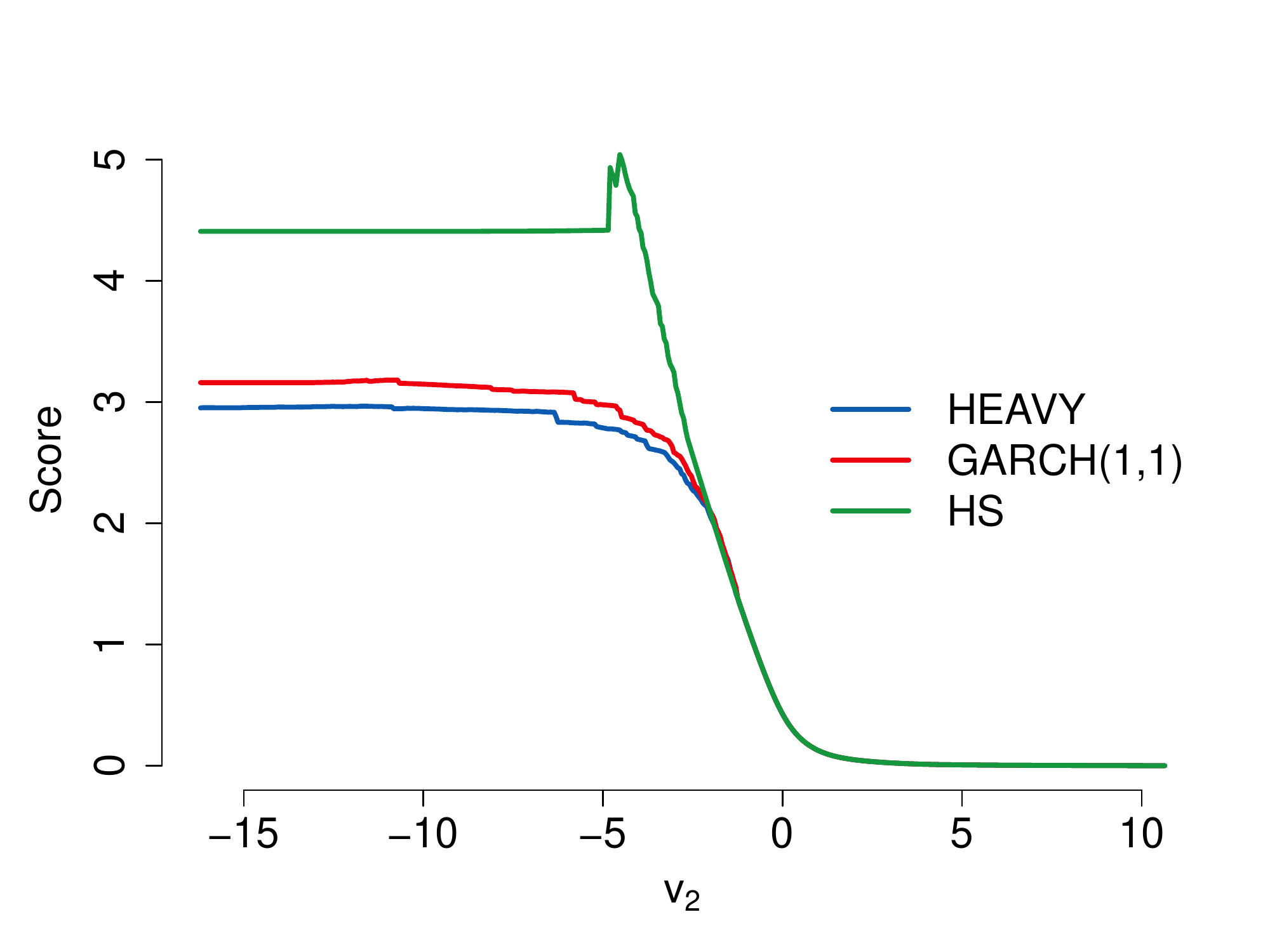}} & \multicolumn{2}{c}{\includegraphics[width = .4\textwidth]{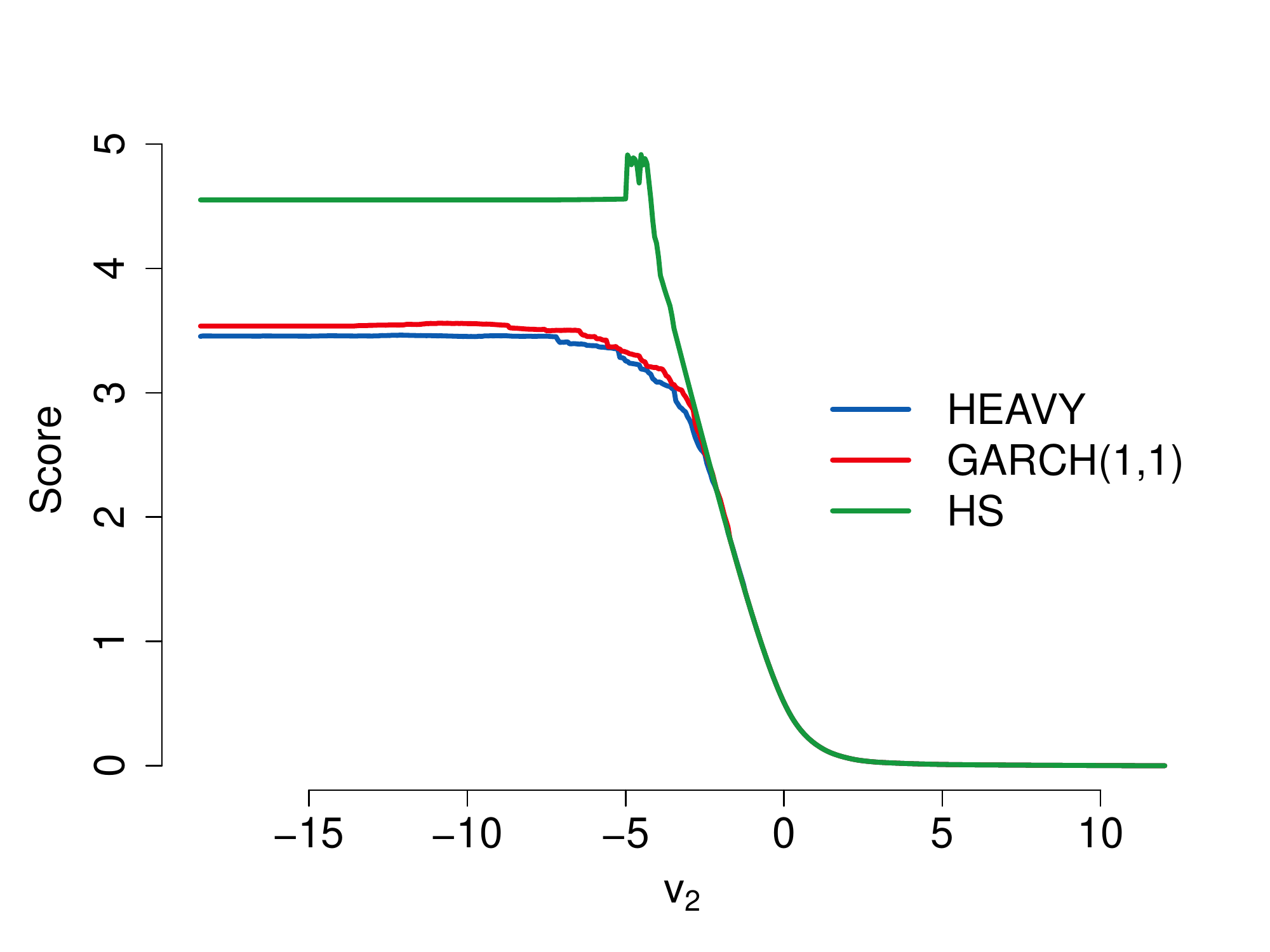}} \\[1ex] \midrule
			&&&\\[-1ex]
			\multicolumn{2}{c}{Difference of HEAVY vs.} & \multicolumn{2}{c}{Difference of HEAVY vs.}\\[1ex]
			HS & GARCH &  HS &  GARCH \\[-3ex]
			\includegraphics[width = 0.23\textwidth]{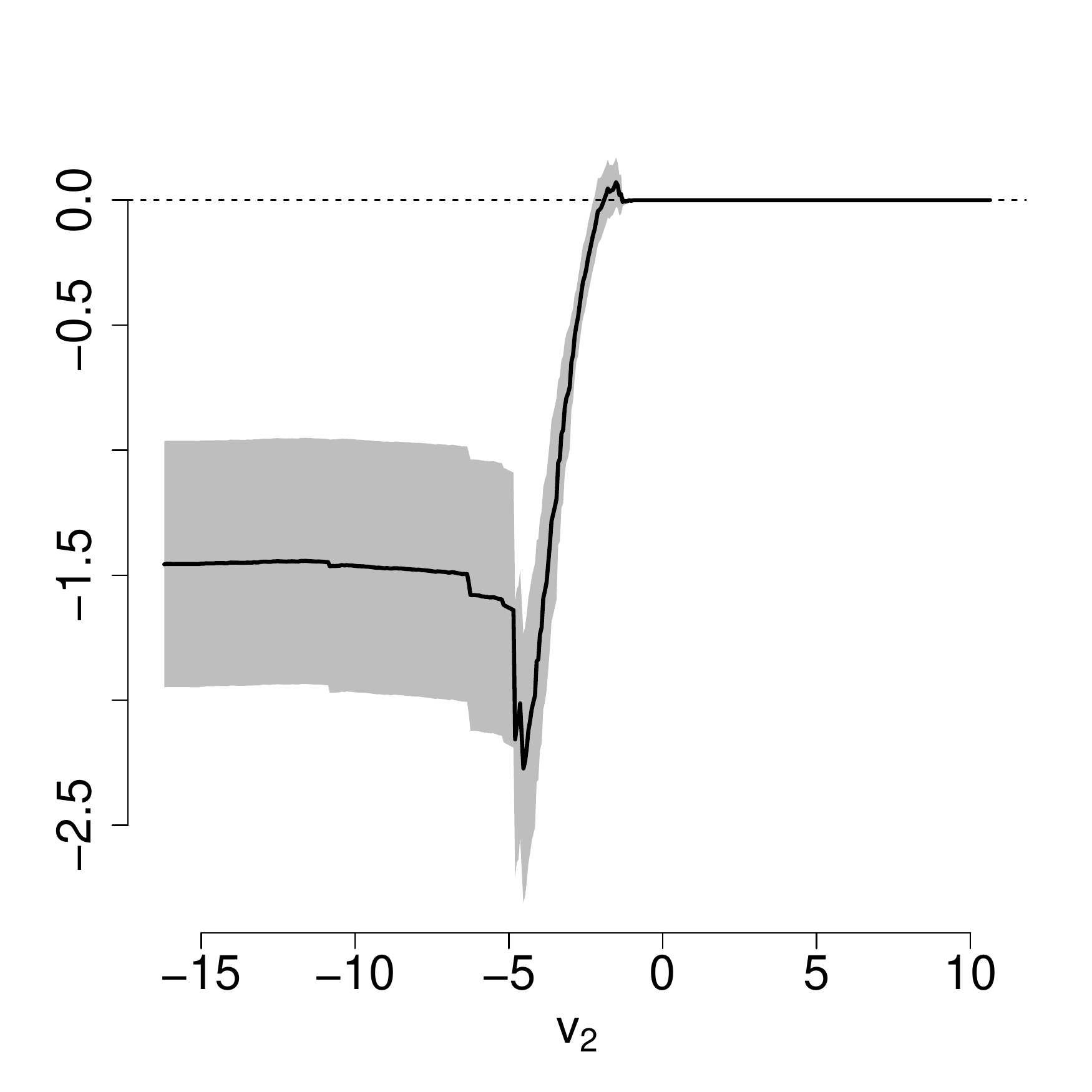} &
			\includegraphics[width = 0.23\textwidth]{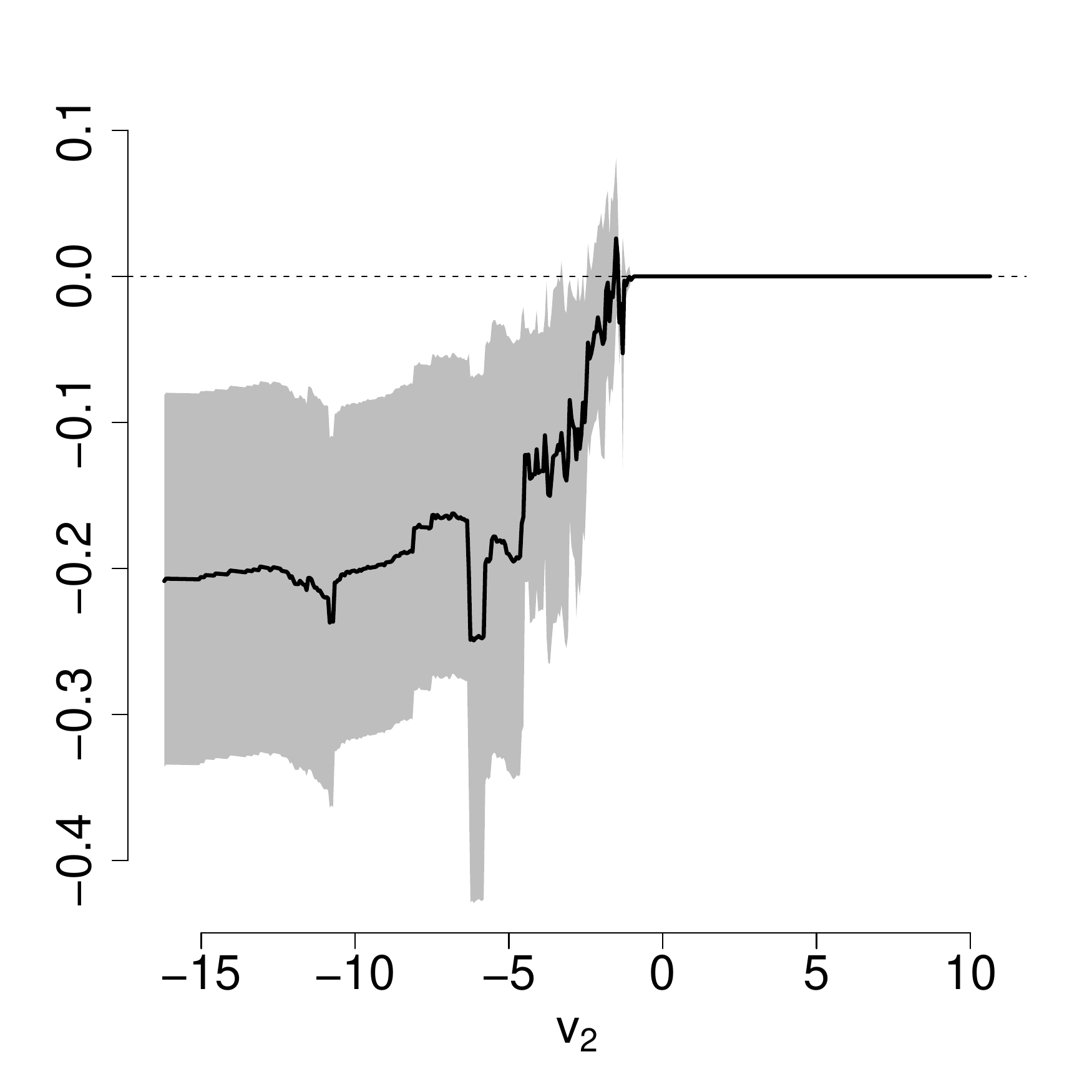} \hspace{.02\textwidth} &
			\hspace{.02\textwidth} \includegraphics[width = 0.23\textwidth]{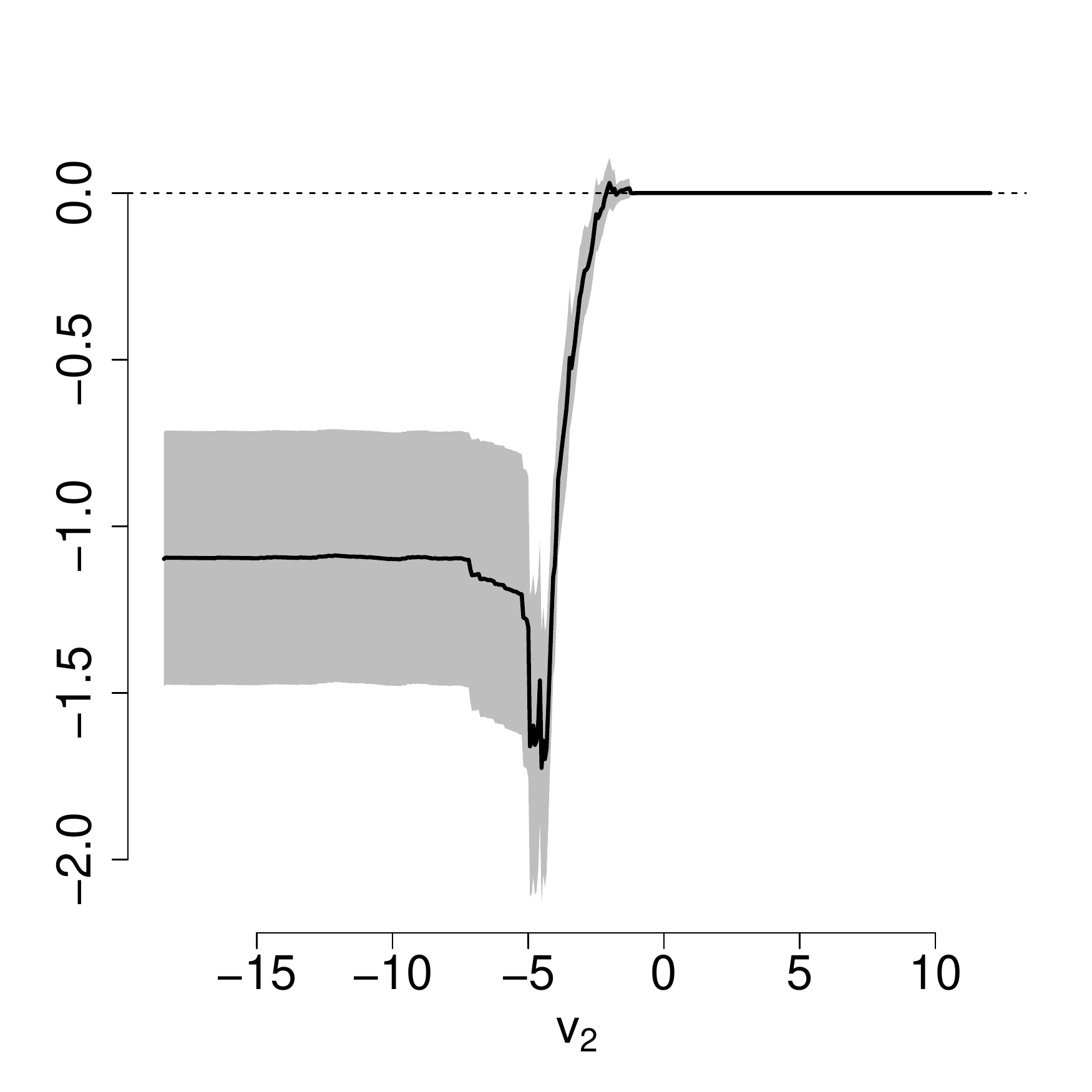} &
			\includegraphics[width = 0.23\textwidth]{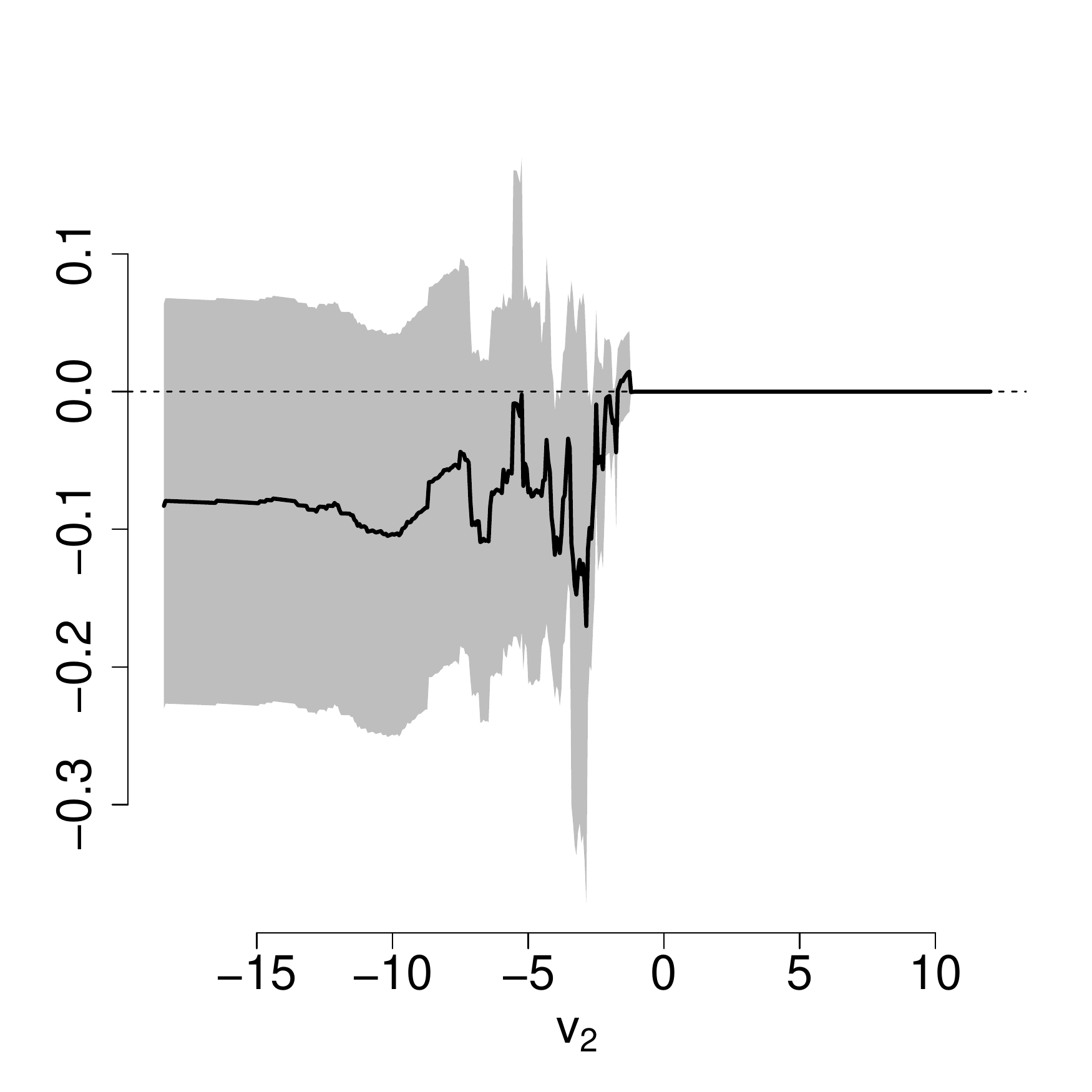} \\ \bottomrule
		\end{tabular} 
\caption{\textbf{Murphy diagrams for empirical forecasts}. Top panels: Smaller scores are better. Bottom panels: Negative difference means that HEAVY outperforms its competitor. Confidence intervals are pointwise at 95\% level.\label{fig:murphy}}	
\end{figure}

\begin{table}[p]
	\begin{tabular}{cc}
		\textbf{\large S\&P 500} & \textbf{\large DAX} \\ \toprule
		\begin{tabular}{lc}
			Hypothesis & P-value \\ \toprule
HS weakly dominates HEAVY & 0.000 \\ 
HEAVY weakly dominates HS & 0.772 \\ [0.3cm]
GARCH weakly dominates HEAVY & 0.000 \\ 
HEAVY weakly dominates GARCH & 0.998 \\ 

		\end{tabular} &
	    \begin{tabular}{lc}
	    	Hypothesis & P-value \\ \toprule
HS weakly dominates HEAVY & 0.000 \\ 
HEAVY weakly dominates HS & 0.876 \\ [0.3cm]
GARCH weakly dominates HEAVY & 0.174 \\ 
HEAVY weakly dominates GARCH & 0.924 \\ 

	    \end{tabular} \\ \bottomrule
    \end{tabular}
\caption{\textbf{Test results for empirical forecasts}. The table presents $p$-values for several hypotheses related to forecast dominance (see Definition \ref{def:dom}). Results are based on the IOT assumption and the class $\mathcal{S}_2$. See Appendix D for additional results.\label{tab:WY}}	
\end{table}

Figure \ref{fig:murphy} and Table \ref{tab:WY} contain the main forecast evaluation results for the S\&P 500 and DAX data sets. We perform forecast evaluation in three steps:
\begin{itemize}
	\item The top row of Figure \ref{fig:murphy} presents Murphy diagrams for all three methods with the display for the S\&P 500 data set at left and the DAX results at right. For both data sets, the HEAVY model seems to attain the lowest average elementary score for the vast majority of thresholds $v_2$. Forecasts based on the GARCH(1,1) model perform slightly worse, and the HS method's performance trails by a considerable margin.
	\item This is emphasized in the bottom row of Figure \ref{fig:murphy}, where the method based on the HEAVY model is compared directly against GARCH(1,1) and HS, respectively. Examining the difference in elementary scores improves our ability to detect which of two models is better at a certain threshold, especially when the difference is small. Pointwise confidence intervals at the 95 percent level (Stage 1 in the dominance test) deliver an impression for the significance of the outperformance exhibited by the HEAVY model. It seems that most examiners would question a significant result only for the comparison of HEAVY to GARCH(1,1) in the DAX data example. However, the final significance decision depends mostly on the way in which the pointwise results are combined.
	\item Table \ref{tab:WY} reports the minimal Westfall-Young p-value of the dominance test: There is ample support against the null hypothesis that HS dominates HEAVY, but no evidence against dominance of HEAVY over HS. These results are found for both the S\&P 500 and the DAX data. In the comparison of HEAVY and GARCH(1,1) for S\&P 500, we similarly find evidence against HEAVY dominating GARCH, but not vice versa. As the previous visual inspection of Figure \ref{fig:murphy} suggests, the HEAVY/GARCH comparison yields different results for the DAX data: At conventional significance levels, we do not find enough evidence to reject either direction of weak dominance.
\end{itemize}
The fact that the HEAVY model tends to outperform its competitors can perhaps be explained by its larger information set, incorporating intra-daily data in addition to daily returns. From \cite{HolzmannEulert2014}, we know that larger information sets lead to better scores \textit{under correct specification}. While the latter assumption is unlikely to be satisfied in practice, one might expect similar results to hold under moderate degrees of misspecification.\\

In Appendix D, we analyze the robustness of our permutation test along two dimensions. First, we compare two different assumptions on the temporal dependence of the elementary scores. Second, we consider using both types of elementary scores $\mathrm{S}_{v_1}$ and $\mathrm{S}_{v_2}$ for the test. The results are generally  similar to the ones reported here, with one exception: When considering both elementary scores without accounting for autocorrelation, the minimal Westfall-Young $p$-values tend to be small for all considered null hypotheses. We conjecture that these results are largely due to non-standard temporal dependence in the first extremal score. However, once one accounts for serial correlation, the results based on both elementary scores are  similar to the ones based on $\mathrm{S}_{v_2}$ only. 

\section{Relationship to Option Pricing}
\label{sec:options}

In Section \ref{sec:2}, we have provided a statistical justification for the class $\mathcal{S}_2$ of scoring functions. We next show that the elementary scores of $\mathcal{S}_2$ also bear an economic interpretation, which resembles connections between VaR, ES and option prices drawn by \cite{Mitra2015} and \cite{BaroneAdesi2016}. Specifically, our elementary score $\mathrm{S}_{v_2}$ is equivalent in decision-theoretic terms to a short position in a European put option with its profit described by
\[
\pi = P - \one\{S \le K\}(K - S),
\]
where $P$ is the put option's price, $K$ is the strike price, and $S$ is the spot price. We draw the elementary scores' relation to $\pi$ by identifying the spot price $S$ with $y$, the strike price $K$ with $x_1$, and by imposing $\alpha (x_1 - v_2)$ as the premium $P$'s structure, such that
\begin{equation} \label{eq:OptionProfit}
\begin{split}
\pi &= \alpha (x_1 - v_2) - \one\{y \le x_1\}(x_1 - y) \\
&= \alpha \one\{v_2 \le y\}(y - v_2) - \alpha \mathrm{S}_{v_2}(x_1, x_2, y),
\end{split}
\end{equation}
conditional on a positive writing decision $x_2 \ge v_2$. Actions are limited to the choice of $x_1$ and $x_2$, corresponding to the strike price and the writing decision, respectively. The first term, $ \alpha \one\{v_2 \le y\}(y - v_2)$, describes the best case scenario without playing a role in the decision-making problem, while the second term can be interpreted as the regret, solely determining the best course of action.\\

Let $F$ denote the distribution of the spot price at maturity of a given asset. From Proposition \ref{prop:mix} and Equation (\ref{eq:OptionProfit}), the expected profit is then
\[
\mathbb{E}(\pi) = \one\{v_2 \le x_2\} \left(x_1(\alpha - F(x_1)) + \int_{-\infty}^{x_1}y \diff F(y) - \alpha v_2\right).
\]
The expression in round brackets as a function of $x_1$ is concave with a maximum at $x_1=\VaR_\alpha(F)$ taking the value $\alpha(\ES_\alpha(F)-v_2)$. Therefore, choosing $x_1=\VaR_\alpha(F)$ is the optimal choice for $x_1$ given a positive writing decision $x_2 \ge v_2$. For $x_2$, any choice such that $(x_2 - v_2)(\ES_{\alpha}(F)-v_2) \ge 0$ is optimal.\\

Assuming that all market participants take only optimal actions, no options with non-zero expected profit will be traded. This implies that the option will only be traded if $v_2 = \ES_\alpha(F)$. This implies for the price of the option that 
\begin{equation} \label{eq:BlackScholes}
P = \alpha(\mathrm{VaR}_\alpha(F) - \mathrm{ES}_\alpha(F)).
\end{equation}

Interestingly, Equation (\ref{eq:BlackScholes}) coincides with the famous \cite{Black1973} pricing formula in one particular case. Specifically, assume that the asset price follows a geometric Brownian motion without trend, such that $F$ is a log-normal distribution with parameters $\mu=\log(y_0) - 0.5~\tau^2~t$ and $\sigma = \tau\sqrt{t}$, where $y_0$ denotes the spot price at present, $\tau$ is the annual volatility,  and $t$ is the time to maturity, where $t=1$ corresponds to one year. Under this form of $F$, Equation (\ref{eq:BlackScholes}) recovers the Black-Scholes formula for the price of a European put option under the additional assumption that the risk-free interest rate is zero \citep[see][Chapter 13]{Hull2008}. The calculations that establish the equivalence are presented in Appendix \ref{app:bs}. Of course, Equation (\ref{eq:BlackScholes}) may yield  different prices than Black-Scholes under other forms for $F$. While we do not take a stance on which form for $F$ -- or, more generally, which option pricing scenario -- is most appropriate, the similarity between statistical incentives (represented by elementary scores) and economic incentives (represented by option payoffs) seems intriguing.

\section{Discussion}

In this paper, we provide a mixture representation for the consistent scoring functions for the pair $(\VaR_\alpha,\ES_\alpha)$. This mixture representation facilitates assessments of whether one sequence of predictions for $(\VaR_\alpha,\ES_\alpha)$ dominates another across a suitable, user-specified class of scoring functions. As we are primarily interested in the comparison of the ES forecasts, we focus on a class that puts as much emphasis on ES as possible. We also demonstrate a general principle for the construction of formal statistical tests for forecast dominance. While the test appears to work well in the simulation and data example, a detailed investigation of its theoretical properties is left for future work.\\

When using Murphy diagrams for comparing forecast performance, it is not necessary to select a specific scoring function prior to forecast evaluation. In the presence of possibly misspecified forecasts and non-nested information sets, this is an advantage as any choice of a particular consistent scoring function induces a preference ordering on all possible sequences of forecasts which is usually difficult or impossible to justify, or, even to describe; see \citet{Patton2016}. On the other hand, Murphy diagrams may lead to inconclusive situations in which neither of the two forecast methods dominates the other. This may be undesirable in contexts of decision making. Ideally, future work should develop a deeper understanding of Murphy diagrams, so that they can not only be used to check for forecast dominance but also guide the decision for a consistent scoring function appropriate for a specific application in case that a total order on forecasting methods is needed.

\bibliographystyle{plainnat}
\bibliography{biblio}

\begin{appendix}

\section*{Appendix}

\section{Proof of Proposition \ref{prop:mix}}

\begin{proof}
The $\F_1$-consistency of $\mathrm{S}_{v_1}$ and $\mathrm{S}_{v_2}$ follows directly from \citet[Corollary 5.5]{FisslerZiegel2015}. This implies the $\F_1$-consistency of $\mathrm{S}$ at \eqref{eq:mix} by a small modification of \citet[Theorem 2]{Gneiting2011}. To see that all scoring functions at \eqref{eq:SVaRES} can be written as at \eqref{eq:mix}, observe that an increasing function $G$ can always be written as
\[
G(x) = \int (\one\{v \le x\} - \one\{v \le z\}) \diff H(v),
\]
where $H$ is a locally finite measure and $z\in \mathbb{R}$. As $G_2 \ge 0$, we can assume that the measure $H_2$ puts finite mass on all intervals of the form $(-\infty,x]$ and choose $z = -\infty$. Finally, $G_2$ is strictly increasing if and only if $H_2$ puts positive mass on all open intervals.  
\end{proof}

\section{Details on the permutation test}

Here we provide implementation details for the permutation test introduced in Section \ref{sec:testing}. In Appendix D, we also apply the test for both types of elementary scoring functions. Therefore, we next describe the most general procedure which involves both elementary scores and thus two grids of parameters (for both $v_1$ and $v_2$). The simpler procedure considered Section \ref{sec:testing} follows easily from the more general variant, by omitting the grid for $v_1$.

\subsubsection*{Stage 1}

The pointwise tests are one-sided t-tests. The results in Sections \ref{sec:testing} and \ref{sec:case} are based on the assumption that the score differences are independent over time (IOT); thus, the variance estimator entering the t-tests does not account for possible autocorrelation. In Appendix D, we present robustness checks using an autocorrelation-consistent \cite{NeweyWest1987} variance estimator, as implemented in the function \texttt{NeweyWest} of the R package \texttt{sandwich} \citep{Zeileis2004}, with a truncation lag of three.\\

\subsubsection*{Stage 2}

Our method for correcting the pointwise p-values follows \cite{WestfallYoung1993} and \cite{CoxLee2008}. We consider a total of $2M$ grid points for $v_1$ and $v_2$, leading to the grids $v_1^{1}, \ldots, v_1^{M}$ and $v_2^{M+1}, \ldots, v_2^{2M}$.\footnote{The description in this paragraph loosely follows \citet[Section 6]{StraehlZiegel2015}.} Let $\sigma$ be the permutation of $\{1, \ldots, 2M\}$ such that $p(v_{\cdot}^{\sigma(1)}) \le \ldots \le p(v_{\cdot}^{\sigma(2M)});$ the subindex $\cdot$ equals either 1 or 2. Consider next two vectors of simulated p-values, $p^*(v_1^{1}), \ldots, p^*(v_1^{M})$ and $p^*(v_2^{M+1}), \ldots, p^*(v_2^{2M})$, generated under the null hypothesis (see below). Define $q_m^* = \text{min}\left\{p^*(v_{\cdot}^{\sigma(s)}): s \ge m\right\}$. For example, $q_1^*$ is the smallest of all simulated p-values, $q_2^*$ is the smallest among the simulated p-values at grid points $v_{\cdot}^{\sigma(2)}, \ldots, v_{\cdot}^{\sigma(2M)}$, and so forth. We simulate $L$ sets of p-values, obtaining values $q^*_{m,l}$ for $1 \le m \le 2M$ and $1 \le l \le L$. The adjusted p-values $r_1, \ldots, r_{2M}$ are finally obtained as 
$$r_m = \frac{1}{L} \sum_{l=1}^L \one\left(q^*_{\sigma^{-1}(m), l} \le p(v_{\cdot}^m)\right).$$
The minimal Westfall-Young p-value of the dominance test is given by $\text{min}_{1 \le m \le 2M} \{r_m\}$. We reject the global null hypothesis if this minimum is smaller than $\alpha$.\\

As suggested above, an important implementation aspect is how to enforce the null hypothesis when simulating the p-values. We do this by randomly permuting the labels of the forecasting methods A and B. Specifically, let $$d_{v_1, t} \equiv \mathrm{S}_{v_1}(X_{t,1}^A, Y_t) - \mathrm{S}_{v_1}(X_{t,1}^B, Y_t)$$ denote the score difference between A and B, at time $t$, for the first elementary score, and $$d_{v_2, t} \equiv \mathrm{S}_{v_2}(X_{t, 1}^A, X_{t, 2}^A, Y_t) - \mathrm{S}_{v_2}(X_{t, 1}^B, X_{t, 2}^B, Y_t)$$ denote the score difference between A and B, at time $t$, for the second elementary score. Under the H0 that A weakly dominates B, it holds that $\mathbb{E}(d_{v_1, t}) \le 0$ and $\mathbb{E}(d_{v_2, t}) \le 0$. At the boundary of the null hypothesis, it holds that $\mathbb{E}(d_{v_1, t}) = 0$ and $\mathbb{E}(d_{v_2, t}) = 0$. We enforce the latter equalities by simulating a sequence $s_t \in \{-1, +1\}, t = 1, \ldots, T,$ and putting 
\begin{eqnarray*}
d_{v_1, t}^* &=& s_t~d_{v_1, t}, \\
d_{v_2, t}^* &=& s_t~d_{v_2, t};
\end{eqnarray*}
note that we use the same sign $s_t$ for all values $v_1, v_2$, thus leaving the correlation structure of the grid points across $v_1, v_2$ intact. We then use the simulated time series $(d_{v_1, t}^*, d_{v_2, t}^*)$ to compute the pointwise p-values $p^*(v_1^{1}), \ldots, p^*(v_1^{M})$ and $p^*(v_2^{M+1}), \ldots, p^*(v_2^{2M})$.\\

In our results in Sections \ref{sec:testing} and \ref{sec:case}, we draw the signs $s_t$ independently across time $t$. This procedure is consistent with the IOT assumption that the score differences are not autocorrelated (see Section \ref{sec:testing}). In Appendix D, we present evidence that simulating the signs in blocks of length four (such that four consecutive periods $t$ are multiplied with the same sign) yields similar test results. 

\section{Additional figures}{\label{app:figs}}

\begin{figure}[!h]
\centering
\begin{tabular}{cc}
\multicolumn{2}{c}{S\&P 500}\\ \midrule
VaR & ES \\ [-0.7cm]
\includegraphics[width = 0.5\textwidth]{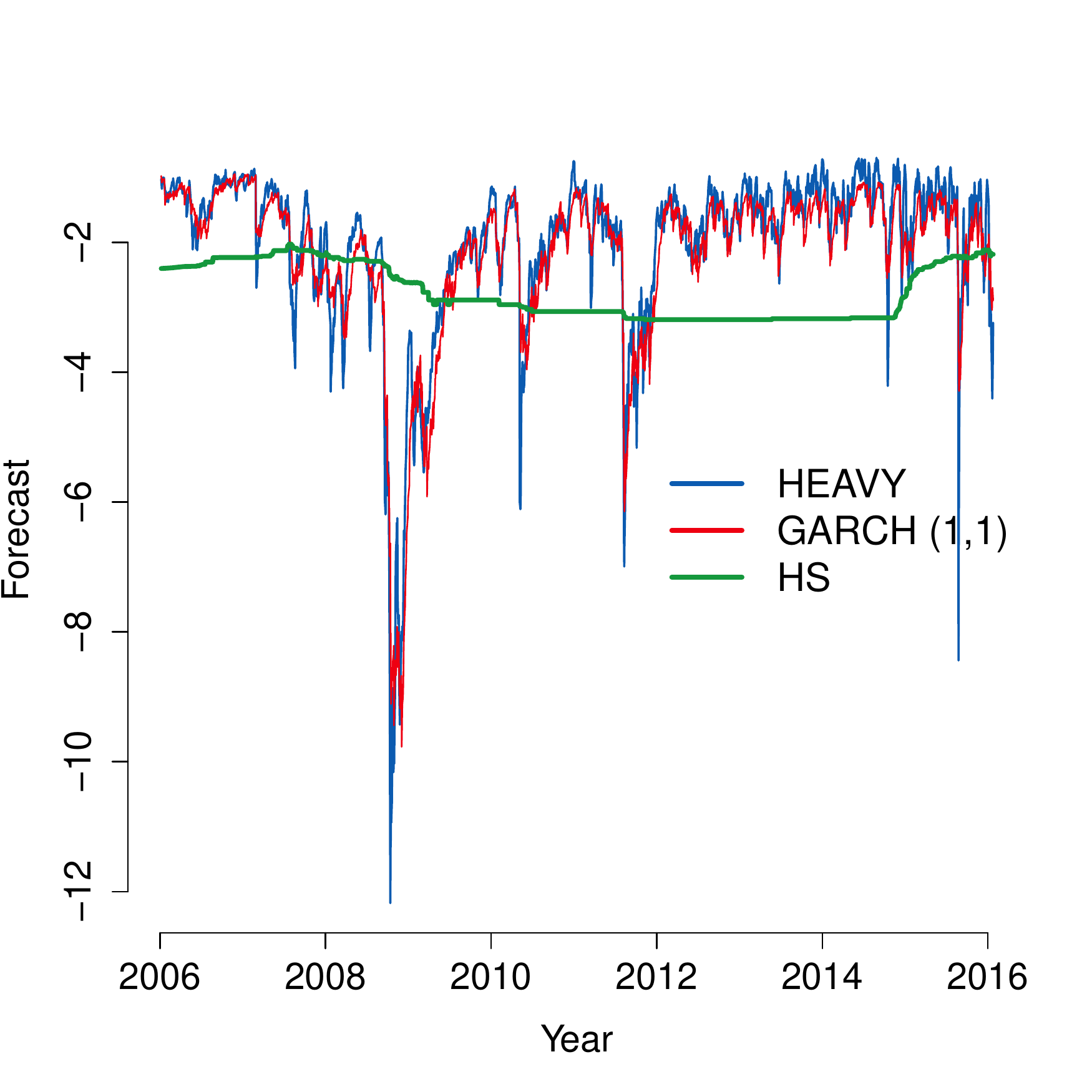} & \includegraphics[width = 0.5\textwidth]{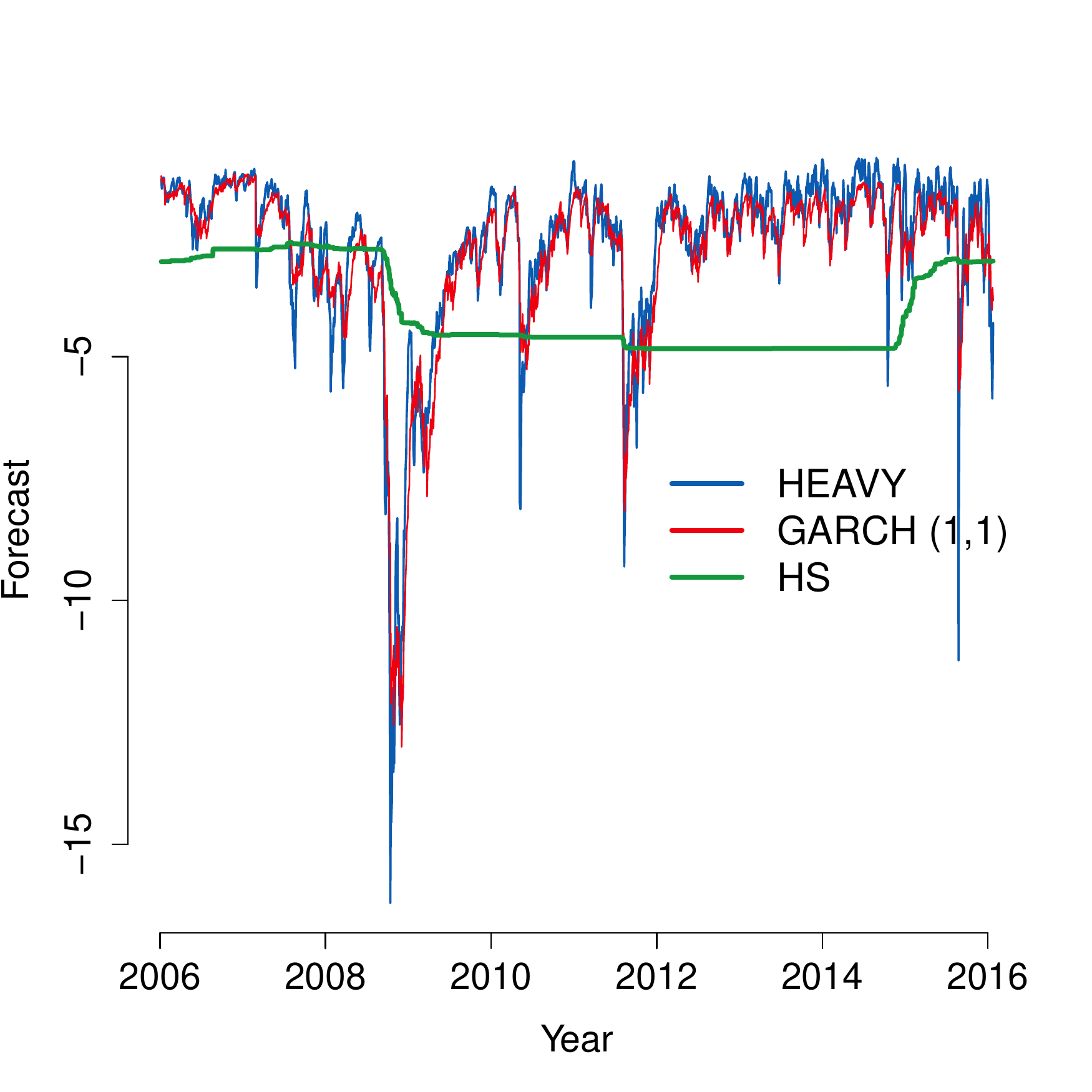} \\ 
\multicolumn{2}{c}{DAX}\\ \midrule
VaR & ES \\ [-0.7cm]
\includegraphics[width = 0.5\textwidth]{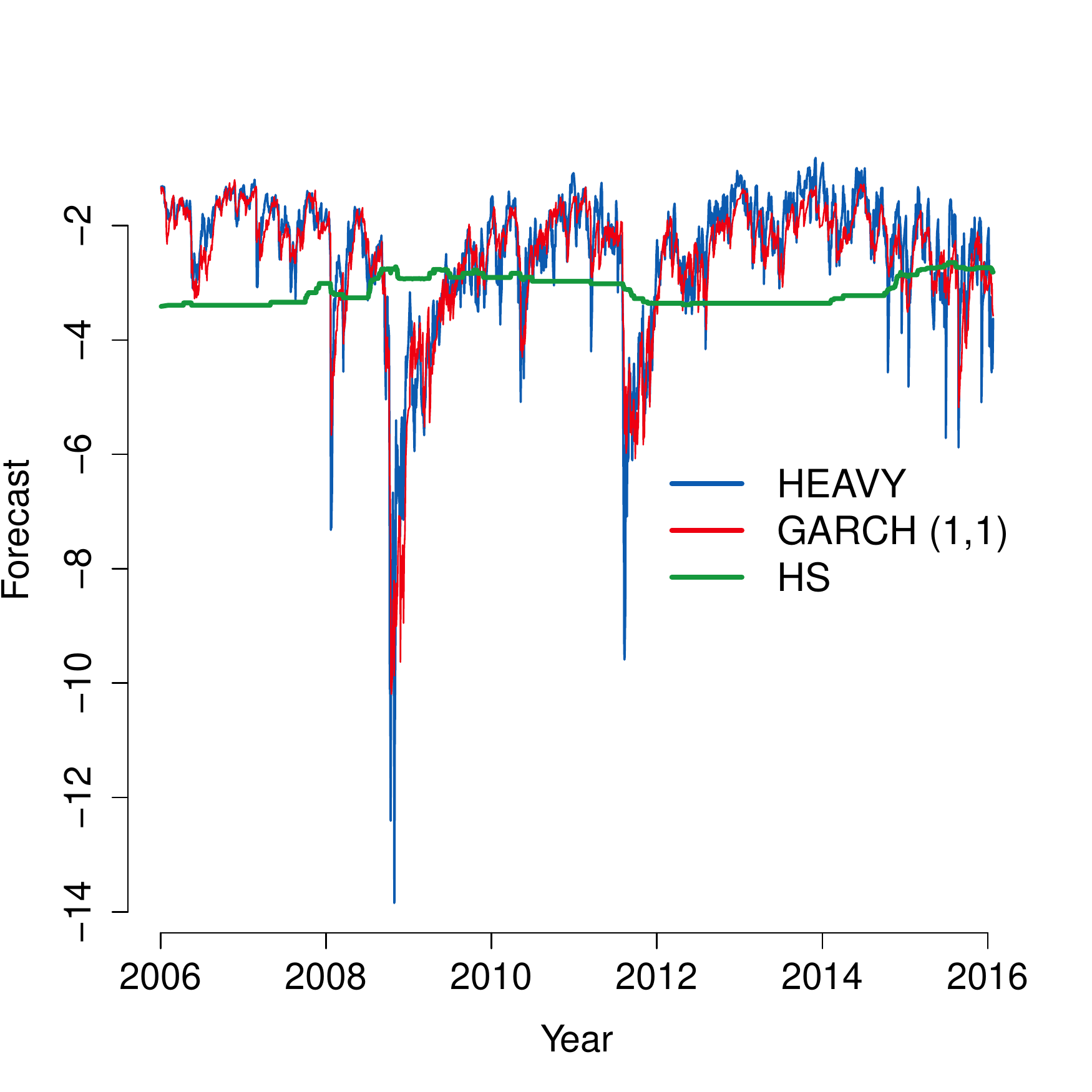} & \includegraphics[width = 0.5\textwidth]{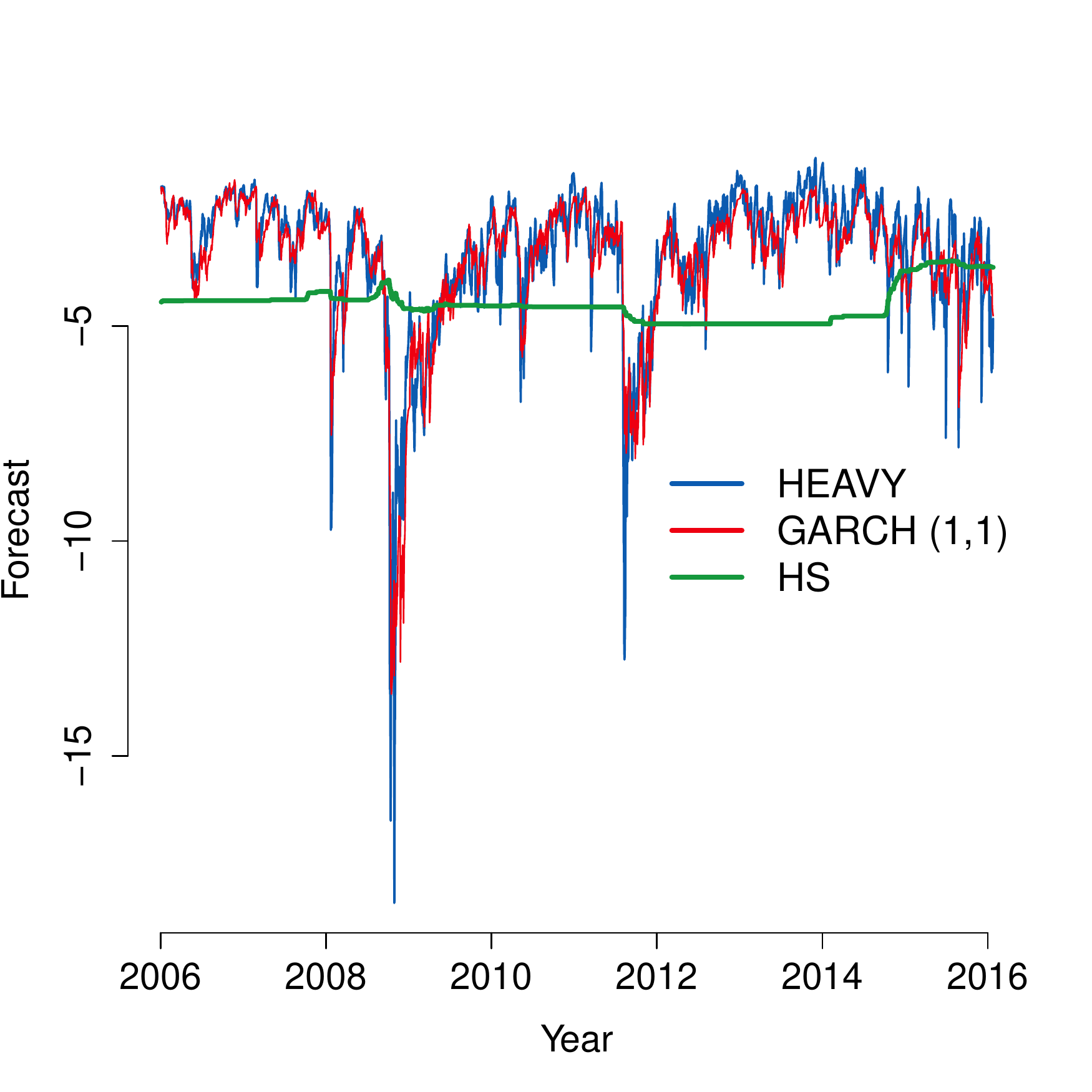} \\
\end{tabular}
\caption{\textbf{Time series plots of empirical forecasts}. Left: Value-at-Risk, right: Expected Shortfall. The sample periods ranges from January 4, 2006 to January 25, 2016. See text for details.}
\label{fig:forecasts}
\end{figure}

\clearpage

\section{Robustness checks for the permutation test}

Here we consider variations of the test described in Section \ref{sec:testing} (and reported in Section \ref{sec:case}), along two dimensions:
\begin{itemize}
	\item \textit{Assumptions on temporal dependence}
	\begin{itemize}
		\item Option 1: Assume temporal independence of the score differences (as in the results in Section \ref{sec:case}). Accordingly, we use independent sign permutations and do not account for possible autocorrelations in the pointwise t-tests.  
		\item Option 2: Allow for temporal independence of the score differences. That is, we use a fixed block length of four when drawing the sign permutations, and a corresponding lag length of three in the pointwise t-tests. 
	\end{itemize} 
	\item \textit{Set of elementary scores considered}
	\begin{itemize}
		\item Option 1: Use all elementary scores.
		\item Option 2: Use only the elementary scores corresponding to the second summand in Equation (3), i.e., only the scores involving both VaR and ES (as in the results in Section \ref{sec:case}). 
	\end{itemize}
\end{itemize}

\begin{table}[!htbp]
\footnotesize	\centering
	\begin{tabular}{cc} \toprule
	\begin{tabular}{lc}
		\multicolumn{2}{c}{\textit{Independence, both elementary scores}} \\ [0.1cm]
		Hypothesis & P-value \\ \toprule
HS weakly dominates HEAVY & 0.000 \\ 
HEAVY weakly dominates HS & 0.008 \\ [0.3cm]
GARCH weakly dominates HEAVY & 0.000 \\ 
HEAVY weakly dominates GARCH & 0.008 \\ 
 
	\end{tabular} & 
\begin{tabular}{lc}
	\multicolumn{2}{c}{\textit{Independence, 2nd elementary score only}} \\ [0.1cm]
	Hypothesis & P-value \\ \toprule
HS weakly dominates HEAVY & 0.000 \\ 
HEAVY weakly dominates HS & 0.772 \\ [0.3cm]
GARCH weakly dominates HEAVY & 0.000 \\ 
HEAVY weakly dominates GARCH & 0.998 \\ 
 
	\end{tabular} \\ [2cm]
\end{tabular} 
\begin{tabular}{cc}
\begin{tabular}{lc}
		\multicolumn{2}{c}{\textit{Dependence, both elementary scores}} \\ [0.1cm]
		Hypothesis & P-value \\ \toprule
HS weakly dominates HEAVY & 0.00 \\ 
HEAVY weakly dominates HS & 0.43 \\ [0.3cm]
GARCH weakly dominates HEAVY & 0.00 \\ 
HEAVY weakly dominates GARCH & 0.35 \\ 

\end{tabular} & 
\begin{tabular}{lc}
		\multicolumn{2}{c}{\textit{Dependence, 2nd elementary score only}} \\ [0.1cm]
		Hypothesis & P-value \\ \toprule
HS weakly dominates HEAVY & 0.00 \\ 
HEAVY weakly dominates HS & 0.79 \\ [0.3cm]
GARCH weakly dominates HEAVY & 0.00 \\ 
HEAVY weakly dominates GARCH & 1.00 \\ 
 
	\end{tabular} \\ 
&\\ \bottomrule
\end{tabular}  
	\caption{\textbf{Test results for empirical forecasts (S\&P 500)}. The table presents several variants of the permutation test, see text for details.}
\end{table}

\begin{table}[!htbp]
\footnotesize	\centering
	\begin{tabular}{cc} \toprule
		\begin{tabular}{lc}
			\multicolumn{2}{c}{\textit{Independence, both elementary scores}} \\ [0.1cm]
			Hypothesis & P-value \\ \toprule
HS weakly dominates HEAVY & 0.000 \\ 
HEAVY weakly dominates HS & 0.000 \\ [0.3cm]
GARCH weakly dominates HEAVY & 0.050 \\ 
HEAVY weakly dominates GARCH & 0.094 \\ 
 
		\end{tabular} & 
		\begin{tabular}{lc}
			\multicolumn{2}{c}{\textit{Independence, 2nd elementary score only}} \\ [0.1cm]
			Hypothesis & P-value \\ \toprule
HS weakly dominates HEAVY & 0.000 \\ 
HEAVY weakly dominates HS & 0.876 \\ [0.3cm]
GARCH weakly dominates HEAVY & 0.174 \\ 
HEAVY weakly dominates GARCH & 0.924 \\ 
 
		\end{tabular} \\ [2cm]
	\end{tabular} 
	\begin{tabular}{cc}
		\begin{tabular}{lc}
			\multicolumn{2}{c}{\textit{Dependence, both elementary scores}} \\ [0.1cm]
			Hypothesis & P-value \\ \toprule
HS weakly dominates HEAVY & 0.000 \\ 
HEAVY weakly dominates HS & 0.072 \\ [0.3cm]
GARCH weakly dominates HEAVY & 0.524 \\ 
HEAVY weakly dominates GARCH & 0.616 \\ 

		\end{tabular} & 
		\begin{tabular}{lc}
			\multicolumn{2}{c}{\textit{Dependence, 2nd elementary score only}} \\ [0.1cm]
			Hypothesis & P-value \\ \toprule
HS weakly dominates HEAVY & 0.000 \\ 
HEAVY weakly dominates HS & 0.846 \\ [0.3cm]
GARCH weakly dominates HEAVY & 0.146 \\ 
HEAVY weakly dominates GARCH & 0.910 \\ 
 
		\end{tabular} \\ 
		&\\ \bottomrule
	\end{tabular}  
	\caption{\textbf{Test results for empirical forecasts (DAX)}. The table presents several variants of the permutation test, see text for details.}
\end{table}

\clearpage

\section{Additional calculations for Section 5}

\label{app:bs}

Here we establish the equivalence between Equation (\ref{eq:BlackScholes}) and the \cite{Black1973} pricing model as noted in Section \ref{sec:options}. To this end, we express Equation (\ref{eq:BlackScholes}) in terms of three factors: The strike price ($x_1$), the current spot price of the underlying asset ($y_0$), and the time to maturity ($t$). We proceed as follows:
\begin{itemize}
	\item In Equation (\ref{eq:BlackScholes}), set $\mathrm{VaR}_\alpha(F) = x_1$. This step follows from the optimality condition described in the text. 
	\item Since $F$ follows a lognormal distribution, we have that $$\alpha = \int_{z = -\infty}^{x_1} d~F(z) = \Phi\left(\frac{\ln x_1 - \ln y_0 + 0.5~\tau^2~t}{\tau \sqrt{t}}\right),$$ where $\Phi$ is the cumulative distribution function of the standard normal distribution. 
	\item Finally, compute $\mathrm{ES}_\alpha(F) = \mathbb{E}(Y|Y < x_1)$. To this end, note that if $Y$ follows a lognormal distribution, then $(\ln Y| \ln Y < \ln x_1)$ follows a truncated normal distribution. Using the moment-generating function of the latter distribution, we obtain
	$$\mathrm{ES}_\alpha(F) = \frac{y_0}{\alpha}~\Phi\left(\frac{\ln x_1 - \ln y_0 - 0.5~\tau^2~t}{\tau \sqrt{t}}\right).$$
	\item Collecting terms, we find that 
	$$P = x_1~\Phi\left(\frac{\ln x_1 - \ln y_0 + 0.5~\tau^2~t}{\tau \sqrt{t}}\right) - y_0~\Phi\left(\frac{\ln x_1 - \ln y_0 - 0.5~\tau^2~t}{\tau \sqrt{t}}\right);$$
\end{itemize}
the latter formula is equal to the Black-Scholes put price in Equation (13.21) of \cite{Hull2008} if the risk-free interest rate is zero.\hfill $\square$

\end{appendix}

\end{document}